\documentclass[preprint,10pt]{elsarticle}

\usepackage{multirow}
\usepackage{amsmath,amssymb,amsfonts}
\usepackage{algorithmic}
\usepackage{lscape}
\usepackage{textcomp}
\usepackage{xcolor}
\usepackage{comment}
\usepackage{geometry}
\usepackage{longtable}
\usepackage{array}
\usepackage[utf8]{inputenc}
\usepackage{subcaption}
\usepackage{graphicx}
\usepackage{rotating} % rotate uml
\usepackage{soul}
\usepackage{longtable}
\usepackage{pgf}
\usepackage{lmodern}
\usepackage{url}
\usepackage{rotating}

\usepackage[acronyms,nonumberlist,nopostdot,nomain,nogroupskip,acronymlists={hidden}]{glossaries}
\newglossary[algh]{hidden}{acrh}{acnh}{Hidden Acronyms}
\newacronym{3gpp}{3GPP}{3rd Generation Partnership Project}
\newacronym{4g}{4G}{4th generation}
\newacronym{5g}{5G}{5th generation}
\newacronym{6g}{6G}{6th generation}
\newacronym{5gc}{5GC}{5G Core}
\newacronym{adc}{ADC}{Analog to Digital Converter}
\newacronym{aerpaw}{AERPAW}{Aerial Experimentation and Research Platform for Advanced Wireless}
\newacronym{ai}{AI}{Artificial Intelligence}
\newacronym{aimd}{AIMD}{Additive Increase Multiplicative Decrease}
\newacronym{am}{AM}{Acknowledged Mode}
\newacronym{amc}{AMC}{Adaptive Modulation and Coding}
\newacronym{amf}{AMF}{Access and Mobility Management Function}
\newacronym{aops}{AOPS}{Adaptive Order Prediction Scheduling}
\newacronym{api}{API}{Application Programming Interface}
\newacronym{xapp}{xApp}{Intelligent Application}
\newacronym{apn}{APN}{Access Point Name}
\newacronym{ap}{AP}{Application Protocol}
\newacronym{aqm}{AQM}{Active Queue Management}
\newacronym{ausf}{AUSF}{Authentication Server Function}
\newacronym{asn1}{ASN.1}{Abstract Syntax Notation One}
\newacronym{avc}{AVC}{Advanced Video Coding}
\newacronym{awgn}{AGWN}{Additive White Gaussian Noise}
\newacronym{balia}{BALIA}{Balanced Link Adaptation Algorithm}
\newacronym{bbu}{BBU}{Base Band Unit}
\newacronym{bdp}{BDP}{Bandwidth-Delay Product}
\newacronym{ber}{BER}{Bit Error Rate}
\newacronym{bf}{BF}{Beamforming}
\newacronym{bler}{BLER}{Block Error Rate}
\newacronym{brr}{BRR}{Bayesian Ridge Regressor}
\newacronym{bs}{BS}{Base Station}
\newacronym{bsr}{BSR}{Buffer Status Report}
\newacronym{bss}{BSS}{Business Support System}
\newacronym{bwp}{BWP}{Bandwidth Part}
\newacronym{ca}{CA}{Carrier Aggregation}
\newacronym{caas}{CaaS}{Connectivity-as-a-Service}
\newacronym{cb}{CB}{Code Block}
\newacronym{cbrs}{CBRS}{Citizens Broadband Radio Service}
\newacronym{cc}{CC}{Congestion Control}
\newacronym{ccid}{CCID}{Congestion Control ID}
\newacronym{cco}{CC}{Carrier Component}
\newacronym{cdd}{CDD}{Cyclic Delay Diversity}
\newacronym{cdf}{CDF}{Cumulative Distribution Function}
\newacronym{cdn}{CDN}{Content Distribution Network}
\newacronym{cn}{CN}{Core Network}
\newacronym{codel}{CoDel}{Controlled Delay Management}
\newacronym{comac}{COMAC}{Converged Multi-Access and Core}
\newacronym{cord}{CORD}{Central Office Re-architected as a Datacenter}
\newacronym{cornet}{CORNET}{COgnitive Radio NETwork}
\newacronym{cosmos}{COSMOS}{Cloud Enhanced Open Software Defined Mobile Wireless Testbed for City-Scale Deployment}
\newacronym{cots}{COTS}{Commercial Off-the-Shelf}
\newacronym{cp}{CP}{Control Plane}
\newacronym{cpu}{CPU}{Central Processing Unit}
\newacronym{cqi}{CQI}{Channel Quality Information}
\newacronym{cr}{CR}{Cognitive Radio}
\newacronym{cran}{CRAN}{Cloud \gls{ran}}
\newacronym{crs}{CRS}{Cell Reference Signal}
\newacronym{csi}{CSI}{Channel State Information}
\newacronym{csirs}{CSI-RS}{Channel State Information - Reference Signal}
\newacronym{cu}{CU}{Central Unit}
\newacronym{d2tcp}{D$^2$TCP}{Deadline-aware Data center TCP}
\newacronym{d3}{D$^3$}{Deadline-Driven Delivery}
\newacronym{dac}{DAC}{Digital to Analog Converter}
\newacronym{dag}{DAG}{Directed Acyclic Graph}
\newacronym{das}{DAS}{Distributed Antenna System}
\newacronym{dash}{DASH}{Dynamic Adaptive Streaming over HTTP}
\newacronym{dc}{DC}{Dual Connectivity}
\newacronym{dccp}{DCCP}{Datagram Congestion Control Protocol}
\newacronym{dce}{DCE}{Direct Code Execution}
\newacronym{dci}{DCI}{Downlink Control Information}
\newacronym{dctcp}{DCTCP}{Data Center TCP}
\newacronym{dl}{DL}{Downlink}
\newacronym{dmr}{DMR}{Deadline Miss Ratio}
\newacronym{dmrs}{DMRS}{DeModulation Reference Signal}
\newacronym{drlcc}{DRL-CC}{Deep Reinforcement Learning Congestion Control}
\newacronym{drs}{DRS}{Discovery Reference Signal}
\newacronym{du}{DU}{Distributed Unit}
\newacronym{e2e}{E2E}{end-to-end}
\newacronym{ecaas}{ECaaS}{Edge-Cloud-as-a-Service}
\newacronym{ecn}{ECN}{Explicit Congestion Notification}
\newacronym{edf}{EDF}{Earliest Deadline First}
\newacronym{embb}{eMBB}{Enhanced Mobile Broadband}
\newacronym{empower}{EMPOWER}{EMpowering transatlantic PlatfOrms for advanced WirEless Research}
\newacronym{enb}{eNB}{evolved Node Base}
\newacronym{endc}{EN-DC}{E-UTRAN-\gls{nr} \gls{dc}}
\newacronym{epc}{EPC}{Evolved Packet Core}
\newacronym{eps}{EPS}{Evolved Packet System}
\newacronym{es}{ES}{Edge Server}
\newacronym{etsi}{ETSI}{European Telecommunications Standards Institute}
\newacronym[firstplural=Estimated Times of Arrival (ETAs)]{eta}{ETA}{Estimated Time of Arrival}
\newacronym{eutran}{E-UTRAN}{Evolved Universal Terrestrial Access Network}
\newacronym{faas}{FaaS}{Function-as-a-Service}
\newacronym{fapi}{FAPI}{Functional Application Platform Interface}
\newacronym{fdd}{FDD}{Frequency Division Duplexing}
\newacronym{fdm}{FDM}{Frequency Division Multiplexing}
\newacronym{fdma}{FDMA}{Frequency Division Multiple Access}
\newacronym{fed4fire}{FED4FIRE+}{Federation 4 Future Internet Research and Experimentation Plus}
\newacronym{fifo}{FIFO}{First In First Out}
\newacronym{fir}{FIR}{Finite Impulse Response}
\newacronym{fit}{FIT}{Future \acrlong{iot}}
\newacronym{fpga}{FPGA}{Field Programmable Gate Array}
\newacronym{fr1}{FR1}{Frequency Range 1}
\newacronym{fr2}{FR2}{Frequency Range 2}
\newacronym{fr3}{FR3}{Frequency Range 3}
\newacronym{fs}{FS}{Fast Switching}
\newacronym{fscc}{FSCC}{Flow Sharing Congestion Control}
\newacronym{ftp}{FTP}{File Transfer Protocol}
\newacronym{fw}{FW}{Flow Window}
\newacronym{ge}{GE}{Gaussian Elimination}
\newacronym{gnb}{gNB}{Next Generation Node Base}
\newacronym{nextg}{NextG}{Next Generation}
\newacronym{gop}{GOP}{Group of Pictures}
\newacronym{gpr}{GPR}{Gaussian Process Regressor}
\newacronym{gpu}{GPU}{Graphics Processing Unit}
\newacronym{gtp}{GTP}{GPRS Tunneling Protocol}
\newacronym{gtpc}{GTP-C}{GPRS Tunnelling Protocol Control Plane}
\newacronym{gtpu}{GTP-U}{GPRS Tunnelling Protocol User Plane}
\newacronym{gtpv2c}{GTPv2-C}{\gls{gtp} v2 - Control}
\newacronym{gw}{GW}{Gateway}
\newacronym{gui}{GUI}{Graphical User Interface}
\newacronym{harq}{HARQ}{Hybrid Automatic Repeat reQuest}
\newacronym{hetnet}{HetNet}{Heterogeneous Network}
\newacronym{hh}{HH}{Hard Handover}
\newacronym{hol}{HOL}{Head-of-Line}
\newacronym{hqf}{HQF}{Highest-quality-first}
\newacronym{hss}{HSS}{Home Subscription Server}
\newacronym{http}{HTTP}{HyperText Transfer Protocol}
\newacronym{ia}{IA}{Initial Access}
\newacronym{iab}{IAB}{Integrated Access and Backhaul}
\newacronym{ic}{IC}{Incident Command}
\newacronym{ietf}{IETF}{Internet Engineering Task Force}
\newacronym{imsi}{IMSI}{International Mobile Subscriber Identity}
\newacronym{imt}{IMT}{International Mobile Telecommunication}
\newacronym{iot}{IoT}{Internet of Things}
\newacronym{ip}{IP}{Internet Protocol}
\newacronym{ipc}{IPC}{Inter-Process Communication}
\newacronym{isac}{ISAC}{Integrated Sensing and Communication}
\newacronym{itu}{ITU}{International Telecommunication Union}
\newacronym{kpi}{KPI}{Key Performance Indicator}
\newacronym{kpm}{KPM}{Key Performance Measurement}
\newacronym{kvm}{KVM}{Kernel-based Virtual Machine}
\newacronym{llc}{LLC}{Lower Layer Control}
\newacronym{los}{LOS}{Line-of-Sight}
\newacronym{lsm}{LSM}{Link-to-System Mapping}
\newacronym{lstm}{LSTM}{Long Short Term Memory}
\newacronym{lte}{LTE}{Long Term Evolution}
\newacronym{lxc}{LXC}{Linux Container}
\newacronym{m2m}{M2M}{Machine to Machine}
\newacronym{mac}{MAC}{Medium Access Control}
\newacronym{manet}{MANET}{Mobile Ad Hoc Network}
\newacronym{mano}{MANO}{Management and Orchestration}
\newacronym{mc}{MC}{Multi-Connectivity}
\newacronym{mcc}{MCC}{Mobile Cloud Computing}
\newacronym{mchem}{MCHEM}{Massive Channel Emulator}
\newacronym{mcs}{MCS}{Modulation and Coding Scheme}
\newacronym{mec}{MEC}{Multi-access Edge Computing}
\newacronym{mec2}{MEC}{Mobile Edge Cloud}
\newacronym{mfc}{MFC}{Mobile Fog Computing}
\newacronym{mgen}{MGEN}{Multi-Generator}
\newacronym{mi}{MI}{Mutual Information}
\newacronym{mib}{MIB}{Master Information Block}
\newacronym{miesm}{MIESM}{Mutual Information Based Effective SINR}
\newacronym{mimo}{MIMO}{Multiple Input, Multiple Output}
\newacronym{ml}{ML}{Machine Learning}
\newacronym{mlr}{MLR}{Maximum-local-rate}
\newacronym[plural=\gls{mme}s,firstplural=Mobility Management Entities (MMEs)]{mme}{MME}{Mobility Management Entity}
\newacronym{mmtc}{mMTC}{Massive Machine-Type Communications}
\newacronym{mmwave}{mmWave}{millimeter wave}
\newacronym{mpdccp}{MP-DCCP}{Multipath Datagram Congestion Control Protocol}
\newacronym{mptcp}{MPTCP}{Multipath TCP}
\newacronym{mr}{MR}{Maximum Rate}
\newacronym{mrdc}{MR-DC}{Multi \gls{rat} \gls{dc}}
\newacronym{mse}{MSE}{Mean Square Error}
\newacronym{mss}{MSS}{Maximum Segment Size}
\newacronym{mt}{MT}{Mobile Termination}
\newacronym{mtd}{MTD}{Machine-Type Device}
\newacronym{mtu}{MTU}{Maximum Transmission Unit}
\newacronym{mumimo}{MU-MIMO}{Multi-user \gls{mimo}}
\newacronym{music}{MUSIC}{Multiple Signal Classification}
\newacronym{mvno}{MVNO}{Mobile Virtual Network Operator}
\newacronym{nalu}{NALU}{Network Abstraction Layer Unit}
\newacronym{ngrg}{nGRG}{next Generation Research Group}
\newacronym{nas}{NAS}{Non-Access Stratum}
\newacronym{nbiot}{NB-IoT}{Narrow Band IoT}
\newacronym{nfv}{NFV}{Network Function Virtualization}
\newacronym{nfvi}{NFVI}{Network Function Virtualization Infrastructure}
\newacronym{nic}{NIC}{Network Interface Card}
\newacronym{nlos}{NLOS}{Non-Line-of-Sight}
\newacronym{now}{NOW}{Non Overlapping Window}
\newacronym{nsm}{NSM}{Network Service Mesh}
\newacronym[type=hidden]{nr}{NR}{New Radio}
\newacronym{nrf}{NRF}{Network Repository Function}
\newacronym{nsa}{NSA}{Non Stand Alone}
\newacronym{nse}{NSE}{Network Slicing Engine}
\newacronym{nssf}{NSSF}{Network Slice Selection Function}
\newacronym{o2i}{O2I}{Outdoor to Indoor}
\newacronym{oai}{OAI}{OpenAirInterface}
\newacronym{oaicn}{OAI-CN}{\gls{oai} \acrlong{cn}}
\newacronym{oairan}{OAI-RAN}{\acrlong{oai} \acrlong{ran}}
\newacronym{oam}{OAM}{Operations, Administration and Maintenance}
\newacronym{ofdm}{OFDM}{Orthogonal Frequency Division Multiplexing}
\newacronym{olia}{OLIA}{Opportunistic Linked Increase Algorithm}
\newacronym{omec}{OMEC}{Open Mobile Evolved Core}
\newacronym{onap}{ONAP}{Open Network Automation Platform}
\newacronym{onf}{ONF}{Open Networking Foundation}
\newacronym{onos}{ONOS}{Open Networking Operating System}
\newacronym{oom}{OOM}{\gls{onap} Operations Manager}
\newacronym{opnfv}{OPNFV}{Open Platform for \gls{nfv}}
\newacronym{orbit}{ORBIT}{Open-Access Research Testbed for Next-Generation Wireless Networks}
\newacronym{os}{OS}{Operating System}
\newacronym{oss}{OSS}{Operations Support System}
\newacronym{pa}{PA}{Position-aware}
\newacronym{pase}{PASE}{Prioritization, Arbitration, and Self-adjusting Endpoints}
\newacronym{pawr}{PAWR}{Platforms for Advanced Wireless Research}
\newacronym{pbch}{PBCH}{Physical Broadcast Channel}
\newacronym{pcef}{PCEF}{Policy and Charging Enforcement Function}
\newacronym{pcfich}{PCFICH}{Physical Control Format Indicator Channel}
\newacronym{pcrf}{PCRF}{Policy and Charging Rules Function}
\newacronym{per}{PER}{Packet Encoding Rule}
\newacronym{pdcch}{PDCCH}{Physical Downlink Control Channel}
\newacronym{pdcp}{PDCP}{Packet Data Convergence Protocol}
\newacronym{pdsch}{PDSCH}{Physical Downlink Shared Channel}
\newacronym{pdu}{PDU}{Packet Data Unit}
\newacronym{pf}{PF}{Proportionally Fair}
\newacronym{pgw}{PGW}{Packet Gateway}
\newacronym{phich}{PHICH}{Physical Hybrid ARQ Indicator Channel}
\newacronym{phy}{PHY}{Physical}
\newacronym{pmch}{PMCH}{Physical Multicast Channel}
\newacronym{pmi}{PMI}{Precoding Matrix Indicators}
\newacronym{powder}{POWDER}{Platform for Open Wireless Data-driven Experimental Research}
\newacronym{ppo}{PPO}{Proximal Policy Optimization}
\newacronym{ppp}{PPP}{Poisson Point Process}
\newacronym{prach}{PRACH}{Physical Random Access Channel}
\newacronym{prb}{PRB}{Physical Resource Block}
\newacronym{psnr}{PSNR}{Peak Signal to Noise Ratio}
\newacronym{pss}{PSS}{Primary Synchronization Signal}
\newacronym{pucch}{PUCCH}{Physical Uplink Control Channel}
\newacronym{pusch}{PUSCH}{Physical Uplink Shared Channel}
\newacronym{qam}{QAM}{Quadrature Amplitude Modulation}
\newacronym{qci}{QCI}{\gls{qos} Class Identifier}
\newacronym{qoe}{QoE}{Quality of Experience}
\newacronym{qos}{QoS}{Quality of Service}
\newacronym{quic}{QUIC}{Quick UDP Internet Connections}
\newacronym{rach}{RACH}{Random Access Channel}
\newacronym{ran}{RAN}{Radio Access Network}
\newacronym[firstplural=Radio Access Technologies (RATs)]{rat}{RAT}{Radio Access Technology}
\newacronym{rcn}{RCN}{Research Coordination Network}
\newacronym{rec}{REC}{Radio Edge Cloud}
\newacronym{ra}{RA}{Resource Allocation}
\newacronym{red}{RED}{Random Early Detection}
\newacronym{renew}{RENEW}{Reconfigurable Eco-system for Next-generation End-to-end Wireless}
\newacronym{rf}{RF}{Radio Frequency}
\newacronym{rfc}{RFC}{Request for Comments}
\newacronym{rfr}{RFR}{Random Forest Regressor}
\newacronym{ric}{RIC}{RAN Intelligent Controller}
\newacronym{rlc}{RLC}{Radio Link Control}
\newacronym{rlf}{RLF}{Radio Link Failure}
\newacronym{rlnc}{RLNC}{Random Linear Network Coding}
\newacronym{rmr}{RMR}{RIC Message Router}
\newacronym{rmse}{RMSE}{Root Mean Squared Error}
\newacronym{rnis}{RNIS}{Radio Network Information Service}
\newacronym{rr}{RR}{Round Robin}
\newacronym{rrc}{RRC}{Radio Resource Control}
\newacronym{rrm}{RRM}{Radio Resource Management}
\newacronym{rru}{RRU}{Remote Radio Unit}
\newacronym{rs}{RS}{Remote Server}
\newacronym{rsrp}{RSRP}{Reference Signal Received Power}
\newacronym{rsrq}{RSRQ}{Reference Signal Received Quality}
\newacronym{rss}{RSS}{Received Signal Strength}
\newacronym{rssi}{RSSI}{Received Signal Strength Indicator}
\newacronym{rtt}{RTT}{Round Trip Time}
\newacronym{ru}{RU}{Radio Unit}
\newacronym{rw}{RW}{Receive Window}
\newacronym{rx}{RX}{Receiver}
\newacronym{s1ap}{S1AP}{S1 Application Protocol}
\newacronym{sa}{SA}{standalone}
\newacronym{sack}{SACK}{Selective Acknowledgment}
\newacronym{sap}{SAP}{Service Access Point}
\newacronym{sc2}{SC2}{Spectrum Collaboration Challenge}
\newacronym{scef}{SCEF}{Service Capability Exposure Function}
\newacronym{sch}{SCH}{Secondary Cell Handover}
\newacronym{scoot}{SCOOT}{Split Cycle Offset Optimization Technique}
\newacronym{sctp}{SCTP}{Stream Control Transmission Protocol}
\newacronym{sdap}{SDAP}{Service Data Adaptation Protocol}
\newacronym{sdk}{SDK}{Software Development Kit}
\newacronym{sdm}{SDM}{Space Division Multiplexing}
\newacronym{sdma}{SDMA}{Spatial Division Multiple Access}
\newacronym{sdn}{SDN}{Software-defined Networking}
\newacronym{sdr}{SDR}{Software-defined Radio}
\newacronym{seba}{SEBA}{SDN-Enabled Broadband Access}
\newacronym{sgsn}{SGSN}{Serving GPRS Support Node}
\newacronym{sgw}{SGW}{Service Gateway}
\newacronym{si}{SI}{Study Item}
\newacronym{sib}{SIB}{Secondary Information Block}
\newacronym{sinr}{SINR}{Signal to Interference plus Noise Ratio}
\newacronym{sip}{SIP}{Session Initiation Protocol}
\newacronym{siso}{SISO}{Single Input, Single Output}
\newacronym{sla}{SLA}{Service Level Agreement}
\newacronym{sm}{SM}{Service Model}
\newacronym{smf}{SMF}{Session Management Function}
\newacronym{smo}{SMO}{Service Management and Orchestration}
\newacronym{sms}{SMS}{Short Message Service}
\newacronym{smsgmsc}{SMS-GMSC}{\gls{sms}-Gateway}
\newacronym{snr}{SNR}{Signal-to-Noise-Ratio}
\newacronym{son}{SON}{Self-Organizing Network}
\newacronym{sptcp}{SPTCP}{Single Path TCP}
\newacronym{srb}{SRB}{Service Radio Bearer}
\newacronym{srn}{SRN}{Standard Radio Node}
\newacronym{srs}{SRS}{Sounding Reference Signal}
\newacronym{ss}{SS}{Synchronization Signal}
\newacronym{sss}{SSS}{Secondary Synchronization Signal}
\newacronym{st}{ST}{Spanning Tree}
\newacronym{svc}{SVC}{Scalable Video Coding}
\newacronym{tb}{TB}{Transport Block}
\newacronym{tcp}{TCP}{Transmission Control Protocol}
\newacronym{tdd}{TDD}{Time Division Duplexing}
\newacronym{tdm}{TDM}{Time Division Multiplexing}
\newacronym{tdma}{TDMA}{Time Division Multiple Access}
\newacronym{tfl}{TfL}{Transport for London}
\newacronym{tfrc}{TFRC}{TCP-Friendly Rate Control}
\newacronym{tft}{TFT}{Traffic Flow Template}
\newacronym{tgen}{TGEN}{Traffic Generator}
\newacronym{tip}{TIP}{Telecom Infra Project}
\newacronym{tm}{TM}{Transparent Mode}
\newacronym{to}{TO}{Telco Operator}
\newacronym{tr}{TR}{Technical Report}
\newacronym{trp}{TRP}{Transmitter Receiver Pair}
\newacronym{ts}{TS}{Technical Specification}
\newacronym{tti}{TTI}{Transmission Time Interval}
\newacronym{ttt}{TTT}{Time-to-Trigger}
\newacronym{tx}{TX}{Transmitter}
\newacronym{uas}{UAS}{Unmanned Aerial System}
\newacronym{uav}{UAV}{Unmanned Aerial Vehicle}
\newacronym{udm}{UDM}{Unified Data Management}
\newacronym{udp}{UDP}{User Datagram Protocol}
\newacronym{udr}{UDR}{Unified Data Repository}
\newacronym{ue}{UE}{User Equipment}
\newacronym{uhd}{UHD}{\gls{usrp} Hardware Driver}
\newacronym{ul}{UL}{Uplink}
\newacronym{um}{UM}{Unacknowledged Mode}
\newacronym{uml}{UML}{Unified Modeling Language}
\newacronym{up}{UP}{User Plane}
\newacronym{upa}{UPA}{Uniform Planar Array}
\newacronym{upf}{UPF}{User Plane Function}
\newacronym{urllc}{URLLC}{Ultra Reliable and Low Latency Communications}
\newacronym{usa}{U.S.}{United States}
\newacronym{usim}{USIM}{Universal Subscriber Identity Module}
\newacronym{usrp}{USRP}{Universal Software Radio Peripheral}
\newacronym{utc}{UTC}{Urban Traffic Control}
\newacronym{vim}{VIM}{Virtualization Infrastructure Manager}
\newacronym{vm}{VM}{Virtual Machine}
\newacronym{vnf}{VNF}{Virtual Network Function}
\newacronym{volte}{VoLTE}{Voice over \gls{lte}}
\newacronym{voltha}{VOLTHA}{Virtual OLT HArdware Abstraction}
\newacronym{vr}{VR}{Virtual Reality}
\newacronym{vran}{vRAN}{Virtualized \gls{ran}}
\newacronym{vss}{VSS}{Video Streaming Server}
\newacronym{wbf}{WBF}{Wired Bias Function}
\newacronym{wf}{WF}{Waterfilling}
\newacronym{wlan}{WLAN}{Wireless Local Area Network}
\newacronym{osm}{OSM}{Open Source \gls{nfv} Management and Orchestration}
\newacronym{pnf}{PNF}{Physical Network Function}
\newacronym{drl}{DRL}{Deep Reinforcement Learning}
\newacronym{rl}{RL}{Reinforcement Learning}
\newacronym{mtc}{MTC}{Machine-type Communications}
\newacronym{osc}{OSC}{O-RAN Software Community}
\newacronym{rc}{RC}{RAN Control}
\newacronym{dqn}{DQN}{Deep Q-Network}
\newacronym{v2x}{V2X}{Vehicle-to-everything}
\newacronym{gbsm}{GBSM}{Geometry-Based Stochastic Model}
\newacronym{gbs}{GBSM}{Geometry-Based Stochastic}
\newacronym{quadriga}{QuaDRiGa}{QUAsi Deterministic RadIo channel GenerAtor}
\newacronym{relu}{ReLU}{Rectified Linear Unit} 
\newacronym{mpc}{MPC}{Multipath Component}
\newacronym{nn}{NN}{Neural Network}
\newacronym{sgd}{SGD}{Stochastic Gradient Descent}
\newacronym{cpi}{CPI}{Conservative Policy Iteration}
\newacronym{trpo}{TRPO}{Trust Region Policy Optimization}
\newacronym{mrat}{multi-RAT}{Multi-Radio Access Technology}
\newacronym{se}{SE}{Spectrum Efficiency}
\newacronym{marl}{MARL}{Multi-Agent \gls{drl}}
\newacronym{noma}{NOMA}{Non-Orthogonal Multiple Access}
\newacronym{td3}{TD3}{Twin Delayed DDPG}
\newacronym{ddpg}{DDPG}{Deep Deterministic Policy Gradient}
\newacronym{esc}{ESC}{Environmental Sensing Capability}
\newacronym{pal}{PAL}{Priority Access License}
\newacronym{gaa}{GAA}{General Authorized Access}
\newacronym{sas}{SAS}{Spectrum Access System}
\newacronym{uci}{UCI}{Uplink Control Information}
\newacronym{lcm}{LCM}{Lifecycle Management}
\newacronym{cnf}{CNF}{Cloud-Native Functions}
\newacronym{nf}{NF}{Network Functions}
\newacronym{dms}{DMS}{Deployment Management Services}
\newacronym{ie}{IE}{Information Element}
\newacronym{sdlc}{SDLC}{Software Development Lifecycle}
\newacronym{lmf}{LMF}{Location Management Function}
\newacronym{cir}{CIR}{Channel Impulse Response}
\newacronym{ci}{CI}{Continuous Integration}
\newacronym{cd}{CD}{Continuous Deployment}
\newacronym{ct}{CT}{Continuous Testing}
\glsdisablehyper

\usepackage{xspace}
\newcommand{\oran}{O-RAN\xspace}
\newcommand{\ran}{\gls{ran}\xspace}

\newcommand{\nearrt}{Near-RT\xspace}
\newcommand{\nonrt}{Non-RT\xspace}
\newcommand{\aiml}{\gls{ai}/\gls{ml}\xspace}
\newcommand{\ric}{\gls{ric}\xspace}
\newcommand{\rics}{\glspl{ric}\xspace}

\newcommand{\ngrg}{\gls{ngrg}\xspace}

\newcommand{\new}[1]{#1}

\graphicspath{{images/}}

\journal{Computer Networks}

\newif\ifexttikz
\exttikzfalse
\ifexttikz
\else
\usepackage{tikzpagenodes,etoolbox}
\usetikzlibrary{calc}
\usepackage[contents={}]{background}
\AddEverypageHook{%
\ifnumequal{\thepage}{1}{%
    \tikz[remember picture,overlay]{%
        \node[draw,
        minimum width=1.03\textwidth,
        text width=1.02\textwidth,
        font=\footnotesize
        ]
        at ($(current page header area)$)
        {%
        This paper has been published in open access on Elsevier Computer Networks. This is the author's accepted version of the article. The final version published by Elsevier is Andrea Lacava, Leonardo Bonati, Niloofar Mohamadi, Rajeev Gangula, Florian Kaltenberger, Pedram Johari, Salvatore D’Oro, Francesca Cuomo, Michele Polese, Tommaso Melodia, ``dApps: Enabling real-time AI-based Open RAN control'', Computer Networks, Volume 269, 2025, 111342, ISSN 1389-1286, https://doi.org/10.1016/j.comnet.2025.111342.
        };
        \node[draw,
        minimum width=1.03\textwidth,
        text width=1.02\textwidth,
        font=\footnotesize
        ]
        at ($(current page footer area)-(0,1cm)$)
        {%
        CC BY-NC 4.0. 
        };
    }%
}{}%end ifnumequal
}
\fi

\begin{document}
\begin{frontmatter}

\title{dApps: Enabling Real-Time AI-Based Open RAN Control}

\author[1,2]{Andrea Lacava\corref{cor1}}
\ead{a.lacava@northeastern.edu}
\author[1]{Leonardo Bonati}
\ead{l.bonati@northeastern.edu}
\author[1]{Niloofar Mohamadi}
\ead{n.mohamadi@northeastern.edu}
\author[1]{Rajeev Gangula}
\ead{r.gangula@northeastern.edu}
\author[1,3]{Florian Kaltenberger}
\ead{f.kaltenberger@northeastern.edu}
\author[1]{Pedram Johari}
\ead{p.johari@northeastern.edu}
\author[1]{Salvatore D'Oro}
\ead{s.doro@northeastern.edu}
\author[2]{Francesca Cuomo}
\ead{francesca.cuomo@uniroma1.it}
\author[1]{Michele Polese}
\ead{m.polese@northeastern.edu}
\author[1]{Tommaso Melodia}
\ead{t.melodia@northeastern.edu}

\address[1]{Institute for the Wireless Internet of Things, Northeastern University, Boston, MA, USA}
\address[2]{Sapienza University of Rome, 00184 Rome, Italy}
\address[3]{EURECOM, Sophia-Antipolis, France}
\cortext[cor1]{Corresponding author}

\begin{abstract}
Open \glspl{ran} leverage disaggregated and programmable \ran functions and open interfaces to enable closed-loop, data-driven radio resource management. This is performed through custom intelligent applications on the \glspl{ric}, optimizing \ran policy scheduling, network slicing, user session management, and medium access control, among others.
In this context, we have proposed dApps as a key extension of the O-\ran architecture into the real-time and user-plane domains.
Deployed directly on \ran nodes, dApps access data otherwise unavailable to \rics due to privacy or timing constraints, enabling the execution of control actions within shorter time intervals.
In this paper, we propose for the first time a reference architecture for dApps, defining their life cycle from deployment by the \gls{smo} to real-time control loop interactions with the \gls{ran} nodes where they are hosted. 
We introduce a new dApp interface, E3, along with an \gls{ap} that supports structured message exchanges and extensible communication for various service models.
By bridging E3 with the existing O-\ran E2 interface, we enable dApps, xApps, and rApps to coexist and coordinate. 
These applications can then collaborate on complex use cases and employ hierarchical control to resolve shared resource conflicts.
Finally, we present and open-source a dApp framework based on \gls{oai}. We benchmark its performance in two real-time control use cases, i.e., spectrum sharing and positioning in a \gls{5g} \gls{gnb} scenario.
Our experimental results show that standardized real-time control loops via dApps are feasible, achieving average control latency below 450\,microseconds and allowing optimal use of shared spectral resources.
\end{abstract}

\glsresetall
\glsunset{nr}

% \begin{highlights} 
% \item The Open RAN ALLIANCE has defined the procedures for enabling control loops for near-real-time and non-real-time domains in the radio resource management of cellular networks, but real-time control loops remain undefined. 
% \item Preliminary work introduced dApps on the DU, enabling access to real-time user-plane data, but without any formalization on how dApps should interact with the rest of the Open RAN architecture. 
% \item We propose and formalize the complete lifecycle of dApps, from deployment to integration within the O-RAN architecture, introducing a new logical interface, E3, which defines formal interactions between dApps, the E2 \ran unit, and xApps.
% \item We present a prototype based on OpenAirInterface and a novel dApp framework that demonstrates the feasibility of dApps through two use cases: Spectrum sharing and sensing and positioning via dApps. The results show that real-time control via dApps can improve the performance of 5G and beyond.
% \item We compare our proposal with competing solutions, particularly the real-time RICs, highlighting their limitations in contrast to our approach which can achieve less than 1\,ms control loops.
% \end{highlights}


\end{frontmatter}

\glsresetall

\section{Introduction}
\label{sec:intro}

% The \oran architecture embraces principles of multi-vendor, softwarized, open and programmable networks toward the next generation of cellular networks.
The Open \gls{ran} architecture promotes open, multi-vendor, software-driven, and programmable cellular networks.
Formalized in the \oran ALLIANCE technical specifications, this novel network architecture introduces the \ric, a software component that leverages open interfaces to gather \ran data, run inference, and enact control, adapting the network to current demands, conditions, and requirements.
%
% The \oran ALLIANCE is coordinating the efforts and releases the specifications tailored at defining a standardization path for Open \ran. 
%
% , and introduces the concept of the \ric. This is a software component that can retrieve data from the \ran using a set of open interfaces, and perform inference or control tasks to tailor the \ran to current user demand, network conditions and requirements.
%
\oran specifications discuss two versions of the \ric. 
% The \ric comes in two embodiments, \nearrt and \nonrt \ric. 
The \nearrt \ric oversees the \ran operations via xApps that operate in the 10\,ms to 1\,s timescale, while the \nonrt \ric hosts rApps that operate at timescales higher than 1\,s.  
%
% The use and combination of xApps and rApps makes it possible to enable a variety of use cases where the \ran can be configured to optimally handle and adapt to network conditions spanning from rapidly changing channel conditions to slowly varying traffic patterns between daytime and nighttime, to name a few.

xApps and rApps enable a variety of use cases where the \ran can be dynamically configured to optimally handle and adapt to varying network conditions.
Wireless systems are characterized by rapidly changing channel conditions, dynamic traffic patterns, user mobility, and periodic or seasonal patterns in network utilization, to name a few.
Examples of \ric-enabled control include network slicing~\cite{johnson2022nexran,polese2022colo}, traffic steering~\cite{lacava2024programmable,dryjanski2021toward}, beamforming and mobility management~\cite{10620740}, advanced sleep modes~\cite{kundu2024towards}, anomaly detection~\cite{bacsaran2023deep}, and spectrum and radio resource allocation~\cite{10356316}.

% Notwithstanding this unprecedented revolution in the \ran control domain, 
The current \oran architecture, though, comes with two key limitations: (i) xApps and rApps are primarily designed to handle control-plane data and operations, thus not considering user-plane data, such as I/Q samples and packets for inference and optimization; 
% \hl{other?} (which carry useful information about channel state and user data)
% underutilized; 
and (ii) does not enable control loops at timescales below the 10\,ms one enabled by xApps.
Indeed, as we will discuss in later sections of this paper, being unable to process user-plane data and perform inference and control below the 10\,ms timescale limits the application of \oran technologies at the lower levels of the protocol stack, as well as the introduction of novel use cases and applications (e.g., \gls{rf} fingerprinting~\cite{al2020exposing}, one-shot beam steering, anomaly and attack detection~\cite{9869746}, spectrum sensing and incumbent detection~\cite{uvaydov2021deepsense}, joint sensing and communications~\cite{chen2008toward,liu2022integrated}, to name a few) which will be at the center of 6G~\cite{qadir2023towards}. 

To overcome these limitations, we proposed the concept of dApps~\cite{d2022dapps}. 
Similarly to xApps and rApps, dApps are software components that can execute as microservices, designed to be co-located with \glspl{cu} and \glspl{du}, where user-plane data is readily available.
Currently, the adoption of dApps is under investigation by the \oran \ngrg as a way to bring below 10\,ms \aiml routines to the \ran. 
% with virtually no data collection latency, as one would instead, have in the case of xApps and rApps retrieving the same data over E2 and O1 interfaces, respectively.
%
The advantages introduced by dApps are manifold: (i) they can execute real-time operations at the \gls{du}/\gls{cu} directly to achieve real-time control and monitoring of the \ran without the need to involve the \glspl{ric}; (ii) they have a lightweight lifecycle management, and can be instantiated and deleted seamlessly to provide functionalities as-a-service based on operator requirements; and (iii) they have access to user-plane data that cannot leave the \gls{du}/\gls{cu} due to privacy concerns (e.g., user-related data) or impracticality (e.g., I/Q streams require high bandwidth in the order of Gbps, which would generate congestion over the other \oran interfaces~\cite{d2022dapps}). 

How to bring \aiml to \oran and push inference toward the real-time domain has received increasing interest in the last few years. Apart from dApps, the literature offers different proposals that include the concepts of real-time \ric and $\mu$Apps~\cite{ko2024edgeric}, zApps~\cite{upadhyaya2023open}, tApps~\cite{liu2023tinyric}, to name a few. 
In this direction, we recently edited an \oran \ngrg research report~\cite{dappsOranReport}, with input and collaborations across academia and industry aimed at exploring use cases and applications that would benefit from dApps.
These include real-time scheduler reconfiguration, spectrum sensing, compute resource scaling for energy savings, channel equalization, beam management and many others. 
% Despite the use cases are clear and well-defined, what is still missing is an architecture to realize dApps and make them viable for cellular applications. 
% Specifically, what is missing is a description of procedures, interfaces, modules and their interactions with already specified \oran components such as xApps, rApps, \rics, \glspl{du}, \glspl{cu} and open interfaces (e.g., O1, O2 and E2).
Despite the well-defined use cases, an architecture for implementing dApps in cellular networks is still lacking. 
In particular, a comprehensive description of the procedures, interfaces, modules, and their interactions with existing \oran components—such as xApps, rApps, \rics, \glspl{du}, \glspl{cu}, and open interfaces (e.g., O1, O2, and E2)—is still missing.

\textbf{Novelty and Contributions.}~In this paper, we fill this gap and propose a reference architecture for dApps with the goal of fostering design and prototyping of dApp-based use cases and applications for \oran systems. Specifically, we propose an architecture that can be seamlessly integrated with the existing \oran architecture with minimal procedural changes. This minimizes the impact that dApps have on \oran standardization, while still making their use feasible and practical. We also propose a \gls{lcm} for the deployment and management of dApps in the \oran architecture.
Once deployed, we show that dApps can leverage already existing interfaces and procedures to exchange data with the \nearrt \ric over the E2 interface and perform monitoring and control tasks using a custom E2\gls{sm} model, i.e., E2\gls{sm}-DAPP. We introduce a novel E3 interface to allow dApps to interact in real time with \glspl{du} and \gls{cu}, for data and control exchange.

We develop and release as open-source\footnote{The framework is available at \url{https://github.com/wineslab/dApp-framework}. A tutorial on how to use the framework is available at \url{https://openrangym.com/tutorials/dapps-oai}.} a framework for dApps integrated with the popular \gls{oai} 5G \gls{ran} framework. 
We evaluate the framework on Colosseum~\cite{polese2024colosseum} and Arena~\cite{BERTIZZOLO2020107436} platforms, conducting an extensive performance analysis to benchmark dApp execution and feasibility. 
Our results demonstrate that dApps operate within real-time control intervals, efficiently processing both vector data (e.g., I/Q arrays extracted from the \ran) and scalar values in loops taking less than 450 microseconds—well below the 10\,ms threshold required for real-time operations.

Additionally, we implement two distinct dApp use cases using the proposed framework \new{for \gls{isac}}: positioning and spectrum sharing.
The results reveal that the positioning dApp is able to compute the distance between the \gls{ue} and the \gls{gnb} using the \gls{ue} \gls{ul} \gls{cir} collected by the \gls{gnb} in real-time, and with the advantage of plug-and-play, customizable processing routines.
Finally, the spectrum-sharing dApp effectively detects incumbents in the \gls{5g} \gls{gnb} \gls{rf} context, enabling more efficient spectrum utilization through real-time analysis.

\new{If compared to~\cite{d2022dapps}, this paper provides a significantly more detailed and practical architecture for dApps integrated with O-RAN systems. Specifically, we provide details on procedures, messages, protocols, and data structures necessary for dApp operation. Additionally, we provide a prototype implementation of dApps in \gls{oai}, demonstrating real-time interactions between dApps and the \gls{ran} protocol stack. Ultimately, we provide an extensive performance evaluation focusing on real-time feasibility and practical benefits. Our results show that dApps can perform inference within $10$\:ms, classify incumbents within $450\:\mu$s, and enable real-time \gls{ue} positioning.}

We believe that the combination of architectural framework, open-source reference library for dApps, and the thorough performance evaluation will inspire further research and development efforts in the real-time control domain.
% , particularly in the spectrum-sharing scenario. 

% While this paper primarily focuses on the prototyping and design of the E3 interface for the DU, the same principles and architecture can be extended to the \gls{cu}. 

% The dApp framework created in the context of this work has been released as an open-source library.

The remainder of this paper is organized as follows.
Section~\ref{sec:soa} presents the current state of the art for the real-time control loops in \oran, highlighting the differences between our work and the real-time \rics proposed in the literature.
Section~\ref{sec:loops} describes the role of the dApps within the \oran architecture and the use cases that can benefit from their implementation.
Section~\ref{sec:arch} presents the integration of the dApps in the \oran architecture, introducing the E3 logic interface and describing the message exchange between the dApps and the other \oran components.
Section~\ref{sec:life} describes the lifecycle of the dApps, from their onboarding through the \gls{smo} to their deployment and in the \ran unit.
In Section~\ref{sec:prototype}, we provide a reference implementation of the architecture proposed based on \gls{oai} and a custom Python framework to measure the real-time control loop communication. We also discuss benchmarking results that profile the performance of the control loop, demonstrating that our real-time control loop can execute in less than $450\,\mu s$.
In Section~\ref{sec:use-cases}, we use our framework to develop two dApp-powered use cases: Spectrum Sharing and Sensing and Positioning.
We evaluate their performance and impact on the \ran, and show how dApps can detect incumbents in a \gls{5g} network and enable spectrum sharing with them.
Section~\ref{sec:conclusions} concludes the paper and envisions future developments for dApps.
Further clarifications on the dApp deployment process are discussed in~\ref{appendix:dapp-deployment}.

% Then we have the near RT control, which is the one that we use for the Radio Resource management use cases such as Traffic Steering and it allows granular control and represents the deepest kind of control you can achieve in \oran at the moment.

% This means that in the state of the art, there are no control loops for what it regards the real-time and the single device, so there is no way to implement intelligence in there.

% At Northeastern, we are designing and pushing as an official proposal to the \oran community a way to achieve such control: the dApps

% The dApps are software programs that use data that cannot be sent out of the \gls{gnb} to implement control policies based on the local conditions of the spectrum and the surrounding environment.
% We can consider them as \gls{ran} functions with steroids, i.e., AI. 

% \hl{lightweight, services, minimal overhead}

% In this work, we extend the definitions of dApps given in~\cite{d2022dapps} and we formalize (propose) a new reference architecture based on this.

% \hl{Real time is mentioned as for further study in the \oran ALLIANCE WG2 report on ``AI/ML workflow description and requirements''}

\section{Related Work and Comparison with Real-time RIC}
\label{sec:soa}

The need for the introduction of real-time control loops in \oran has been advocated in recent works to control lower-layer functionalities of the \gls{gnb} protocol stack~\cite{polese2023understanding,abdalla2022toward}.
In our previous work~\cite{d2022dapps}, we introduced the concept of dApps to process data locally available at the \gls{gnb}, and the need for sub-$10$\:ms control loops.
However, our \new{prior} work primarily focused on \new{advocating} the benefits of such tighter control loops for generic use cases. \new{In this work, instead, we propose a complete reference architecture and design for dApps, including interfaces for interoperability of dApps with the rest of the O-RAN components. Then, we showcase the capabilities and potential of dApps for use cases of interest to Open \gls{ran} networks.}

\begin{figure}[!ht]
\centering
\begin{subfigure}[t]{0.48\columnwidth}
    \centering
    \includegraphics[height=4.5cm]{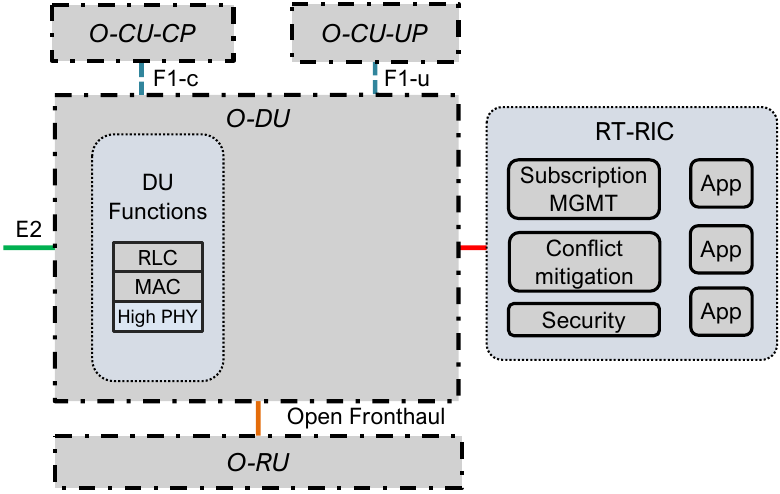}
    \caption{RT-RIC deployment}
\end{subfigure}
\begin{subfigure}[t]{0.48\columnwidth}
    \centering
        \includegraphics[height=4.5cm]{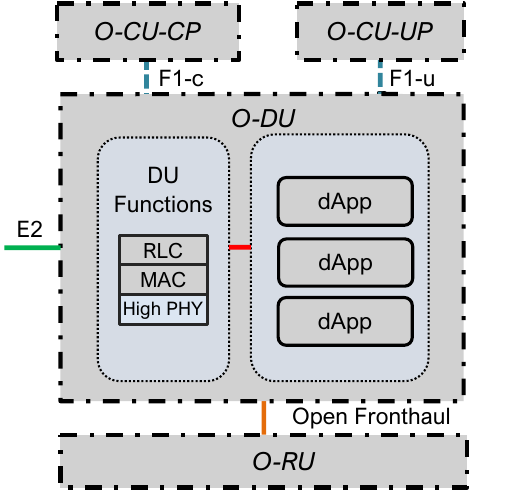}
    \caption{dApp deployment}
 \end{subfigure}
    \caption{Comparison between RT RIC and dApp architectures. From~\cite{dappsOranReport}.}
    \label{fig:rt-ric-dapp-architecture}
\end{figure}

The need for real-time control loops and inference in \oran has also been \new{highlighted} by the broader community. For example, the concept of Real-time \glspl{ric} and its applications have been proposed in~\cite{upadhyaya2023open,liu2023tinyric,ko2024edgeric}.
Specifically,~\cite{upadhyaya2023open} proposes to embed \glspl{cu} and \glspl{du} inside the real-time \gls{ric}, as well as to further disaggregate their functionalities, where each one (e.g., \gls{mac} scheduler, channel estimators, and beam shapers) is carried out by a dedicated application, namely zApp.
\cite{liu2023tinyric} proposes TinyRIC, which runs close to the \gls{cu}/\gls{du}, and so-called tApps hosted therein. The latter aid the \gls{cu}/\gls{du} by performing functionalities spanning from the management of non-\gls{3gpp} interfaces (e.g., O1 and E2), to energy management, to providing primitives for data collection.
Similarly,~\cite{ko2024edgeric} proposed EdgeRIC and its applications, $\mu$Apps, that interface with the \gls{ran} to make decisions at \gls{tti} level.
Although these approaches propose a viable way to introduce real-time control in the \gls{ran}, they do so by replicating functionalities typical of non- and \nearrt \glspl{ric}, e.g., intelligence orchestration, in a new RT \gls{ric} platform, which inherently increases the network complexity, as well as resource utilization and energy consumption.
\new{Our approach, instead, leverages existing \gls{ric} and well-established routines to deploy and operate applications on the \gls{ran} elements, while avoiding the additional complexity and resource utilization that the additional RT controllers would require.}
Indeed, the main difference between the above-mentioned \new{solutions} and dApps lies in the way intelligent applications are hosted and executed.
A high-level architectural comparison of the two approaches is shown in Fig.~\ref{fig:rt-ric-dapp-architecture}, which highlights the placement of intelligent applications in the each of them.
\new{While the above-mentioned applications require a dedicated RT \gls{ric} entity, as well as additional layers of network abstractions, dApps execute at the \gls{cu}/\gls{du} directly. This minimizes the impact on the E2 nodes and takes advantage of interfaces and functionalities already defined in the \oran specifications.}
% \hl{MP: Add comparison diagrams from \oran RR, and maybe we can call this section related work and comparison with RT RIC}
% %
% However, these approach require the introduction of an additional real-time \gls{ric} platform, which would increase the network complexity. Furthermore, some of these approaches would also require the disaggregation of the \gls{cu}/\gls{du} protocol stack layers, with the consequent complete re-design of the 5G protocol stack.
% %
% By leveraging the components specified by \oran, instead, our approach enables real-time control-loops, while keeping the additional network complexity and overhead at bay, as well as limiting the required modifications to existing \oran components.

Finally,~\cite{foukas2023taking} proposes Janus, a monitoring and control framework that can load and execute custom real-time ``codelets'' on the \gls{cu} and \gls{du}. Even though this work---possibly the most similar to ours---extends O-\ran to real-time control, Janus's codelets \new{require direct access to the \gls{cu}/\gls{du} protocol stack rather than interfacing with it, as dApps instead do. In some cases, this can be a limitation as vendors might need to host third-party codelets rather than exposing parameters via interfaces to regulate access to \gls{cu}/\gls{du} functionalities.}
% In addition, the current version of Janus only works with CPU-based stacks.
% , and can only be applied to CPU-based protocol stack implementations.
Moreover, this approach also requires the definition and deployment of a dedicated controller on the \gls{ric}, instead of relying on standardized components for the management of real-time \gls{ran} functions, as we instead do in our dApp architecture.

\new{Table~\ref{tab:dapp_vs_rt_ric} summarizes the key differences between dApps and other real-time control solutions and the advantages of our proposal.
}

\begin{table}[ht!]
    \centering
    \footnotesize
    \renewcommand{\arraystretch}{1.2} % Adjust row spacing
    \setlength{\tabcolsep}{6pt} % Reduce column spacing
    \begin{tabular}{|p{3.2cm}|p{3.3cm}|p{3.3cm}|p{3.3cm}|}
        \hline
        \textbf{Feature} & \textbf{dApps (this work)} & \textbf{RT RIC-based Solutions (zApps~\cite{upadhyaya2023open}, tApps~\cite{liu2023tinyric}, $\mu$Apps~\cite{ko2024edgeric})} & \textbf{Janus}~\cite{foukas2023taking} \\
        \hline
        \textbf{Connectivity to RAN stack} & Disaggregated & Disaggregated & Integrated \\
        \hline
        \textbf{Timescale} & $< 1$\,ms & from $< 1$\,ms ($\mu$App, tApp) to $< 10$\,ms (zApp) & $< 1$\,ms \\
        \hline
        \textbf{Deployment Location} & Directly on CU/DU & Require dedicated RT RIC platform & Directly modifies CU/DU protocol stack \\
        \hline
        \textbf{Additional Network Complexity} & Minimal, leverages existing RIC components & High, add new RIC layers & High, modifies CU/DU stack \\
        \hline
        \textbf{Resource Utilization} & Optimized, avoids additional overhead & Increased due to extra control layers & High, requires additional controllers \\
        \hline
        \textbf{Integration with O-RAN} & Use standardized interfaces and components & Require additional network abstractions & Limited, modifies protocol stack instead of interfacing with it \\
        \hline
        \textbf{Scalability} & High, built on existing RAN elements & Limited, require new infrastructure & Limited, protocol stack modifications reduce compatibility \\
        \hline
        \textbf{Modularity} & High, they interface with CU/DU without modifying stack & Medium, dependent on RT RIC architecture & Low, tightly coupled with CU/DU stack \\
        \hline
        \textbf{Flexibility} & Adapt to different CU/DU implementations & Constrained by RT RIC architecture & Low, as it relies on direct modifications to protocol stack \\
        \hline
        \textbf{Use of Standard O-RAN APIs} & Yes, integrates with O-RAN standard components & Partially, introduce custom abstractions & No, modifies stack implementation directly \\
        \hline
    \end{tabular}
    \caption{Comparison of dApps with RT RIC-based solutions and Janus.}
    \label{tab:dapp_vs_rt_ric}
\end{table}
\newpage

\section{The Role of dApps in the Hierarchical \oran Control Architecture}
\label{sec:loops}

\begin{figure}[ht]
    \centering
    \includegraphics[width=0.95\linewidth]{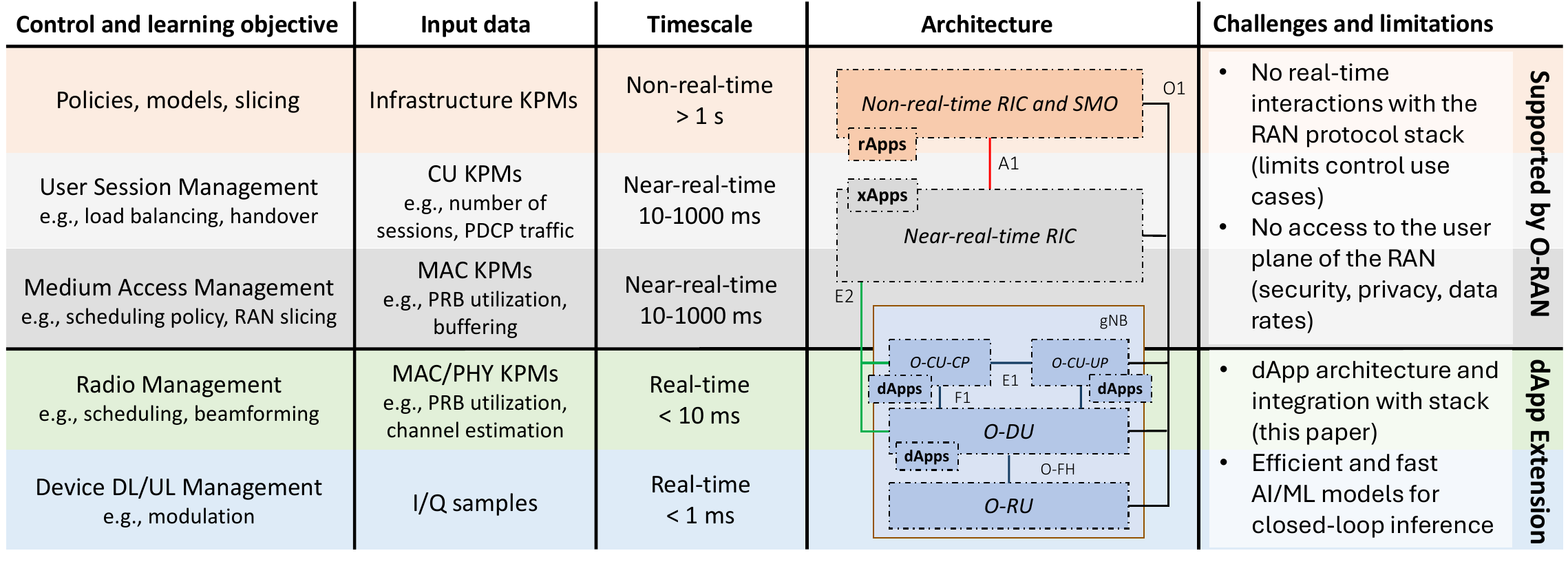}
    \caption{\oran control loops, limitations, and extensions to the real-time and user-plane domains. Adapted from~\cite{polese2023understanding}.}
    \label{fig:loops}
\end{figure}

Achieving programmability of the cellular protocol stack via closed-loop control is a fundamental innovation introduced by the \oran architecture and, more in general, by the Open \gls{ran} vision~\cite{polese2023understanding}. With closed-loop control, data-driven logic units executed in
% custom applications on the \glspl{ric}, i.e., the
xApp and rApps access \gls{ran} telemetry and process it to identify \gls{ran} configurations that satisfy specific intents and objectives of the operator. The configuration is then applied to the \gls{ran} nodes either in form of control (e.g., update a certain \ran parameter) or as a policy, allowing the system to reach the desired state. 

Figure~\ref{fig:loops} reports a simplified view of the \oran logical architecture and control loops. This includes the \gls{ran}, with the Open \glspl{cu}-\gls{cp} and \gls{up}, the \gls{du}, and the \gls{ru}.\footnote{For the purposes of this paper, we consistently refer to the Open RAN units while omitting the prefix ``O-''.} The F1 interface connects CU and DU, while the fronthaul interface bridges the DU and the RU. The E1 interface is in between the \gls{cu}-\gls{cp} and \gls{up} network functions. The \oran architecture leverages the \nonrt \gls{ric} and the \gls{smo} framework for orchestration and policy definition, through the O1 interface to all \gls{ran} nodes and the A1 interface to the \nearrt \glspl{ric}. The latter performs radio resource management with direct control of the \gls{ran} nodes through the E2 interface. These elements are deployed over a set of physical and virtualized resources on a heterogeneous infrastructure (i.e., the O-Cloud). The O2 interface from the \gls{smo} interacts with the O-Cloud to configure the compute layer.

The current \oran architecture features two control loops, at the non-real-time and the near-real-time scales. The first, exercised by the \nonrt \gls{ric}, identifies policies that apply to the network considering \glspl{kpm} reported on a scale of thousands of end devices. It operates at a scale of 1\,s or more. The second loop runs between 10\,ms and 1\,s, and is exercised by the \nearrt \gls{ric}. It can perform control or deploy policies, potentially based on the ones defined by non-real-time loops. Due to the tighter requirements on the completion of the control loop, the \nearrt \gls{ric} usually targets tens of base stations with hundreds of end devices~\cite{maxenti2024scaloran}. 

These loops, however, share some limitations, as discussed in the right portion of Fig.~\ref{fig:loops}, such as the lack of control loops faster than 10\,ms, and the lack of interaction and programmability on the user-plane data. Real-time interactions open up many fine-grained and new customizable and programmable inference and control loop capabilities, including beam management, scheduling profile selection, packet tagging, dynamic spectrum access, and \gls{qos} enforcement, among others. Further, access to user-plane data would enable direct interaction with waveforms and \glspl{pdu} at different layer of the stack, as well as inference and control based on the rich information that these elements carry. There is a significant body of research demonstrating the benefit of inference and classification based on raw I/Q samples for anomaly detection, spectrum sensing, fingerprinting, and beam management, among others~\cite{charm,uvaydov2021deepsense,villa2023twinning,al2020exposing,polese2021deepbeam}. The same applies to user-plane units at higher layers, i.e., transport blocks, \gls{rlc} and \gls{pdcp} packets, e.g., for traffic classification and slice identification~\cite{tractor,johnson2022nexran}. 

However, \oran xApp/rApp-based control loops are not suitable for real-time control. 
For example, transferring I/Q samples or, in general, user data out of the \gls{ran}, is generally unfeasible due to security, privacy, and timing/bandwidth constraints, as also mentioned in Sec~\ref{sec:intro}. 
We discussed this in~\cite{d2022dapps,dappsOranReport}, estimating that transferring I/Q samples through a rate-limited E2 interface would take seconds, which is incompatible with real-time interactions, \new{and can limit the capabilities of the E2\gls{sm} for the \gls{llc}~\cite{oran-wg3-e2-sm-llc}}.
Bringing programmability and observability to \gls{ran} nodes is, thus, paramount to address both limitations. The bottom part of Fig.~\ref{fig:loops} illustrates the role that dApps
% a third category of programmable applications, i.e., the dApps,
can play in future iterations of the \oran architecture. dApps are lightweight and plug-and-play services that provide secure and real-time access to the stack without introducing the overhead that an additional real-time \gls{ric} would bring to network-oriented \gls{ran} nodes. 

\paragraph{Use Cases for dApps and Real-time Control Loops}
In~\cite{dappsOranReport}, recently published as contributed research report by the \oran ALLIANCE, we reviewed how dApps can complete the hierarchical structure of control loops within the \oran architecture. The use cases can be classified into four main groups, described next.

\begin{itemize}
    \item \textbf{Direct Processing of Waveforms and \glspl{pdu}.} Access to I/Q samples can be used for physical layer security, e.g., through anomaly detection to detect spoofing of base stations or legitimate users~\cite{9869746}, and with \gls{rf} fingerprinting for secure authentication. Similarly, information carried by I/Q samples on scheduled or unscheduled symbols can be used to detect spectrum holes or incumbents. This enables spectrum sharing but also remote interference detection, including from other base stations in the operator network. In both cases, information processed at the dApp level can be shared with the stack or xApps to coordinate responses across multiple \gls{ran} nodes, \new{paving the way for the \gls{ran} to function as an \gls{isac} system.}

    \item \textbf{Real-Time Scheduling and Beam Management.} This includes real-time scheduling reconfiguration to embed ad hoc policies and \glspl{sla} for specific traffic patterns and profiles (e.g., new slices), scheduling acceleration for efficient massive \gls{mimo} and multi-cell coordination~\cite{m3mimoJsacPaper}, and real-time sharing coordination with arbitration across different entities~\cite{qualcommSharing}. Similarly, beam management procedures can be enhanced and aided by direct access and manipulation of reference and synchronization signals in the \gls{du}~\cite{giordani2018tutorial}. \new{In this context, dApps can extract real-time measurements, perform data-driven inference at near-real-time periodicity, and send the resulting data to an xApp, which uses it to control beamforming via the existing E2\gls{sm} \gls{llc}~\cite{oran-wg3-e2-sm-llc}.}

    \item \textbf{\gls{ran} Nodes and Fronthaul Configuration.} dApps can be used to dynamically reconfigure and tune the \gls{ran} nodes underlying compute and the fronthaul interface between the \gls{du} and \gls{ru}. For example, dApps can bridge real-time \gls{ran} telemetry from the stack with the compute node configuration to dynamically modify CPU pinning or CPU energy states for energy-efficiency optimization. Similarly, the real-time telemetry can dynamically affect how the fronthaul interface is configured, e.g., adapting the compression level to save resources when users have good \gls{snr} and providing uncompressed streaming for users at the edge of the cell.

    \item \textbf{Augmented Sensing and Channel Estimation.} dApps can enable dynamic \gls{csi} compression in coordination with \glspl{ue}~\cite{haiPaper}, as well as custom \aiml models for channel estimation that can be tailored to specific user conditions and requirements. dApps can also be used to process reference and synchronization signals for use cases not traditionally considered within \gls{ran} nodes, e.g., sensing and positioning~\cite{palama2022}.
\end{itemize}

\section{dApp Service-based Architecture and Integration with RAN Nodes}
\label{sec:arch}

% Here we describe the dApps and how it fits within the \oran architecture. We show dApps, xApps, rApps, and interfaces used to connect them and gather data. Data shared access. Concurrency and coexistence of dApps
% Here we discuss how dApps are integrated with \gls{ran} nodes (O-CU-CP, O-CU-UP, O-DU), from an architectural point of view 
% Provide some high-level discussion on the interface between \gls{ran} nodes and dApps, to be expanded later 
% Discuss how multiple dApps can be supported in this architecture 
% Discuss how a dynamic lifecycle can be supported in this architecture (more details later) 

% Architecture

Based on the use cases and related requirements discussed in Sec.~\ref{sec:loops}, in the following paragraphs we discuss how the \oran architecture can be extended to support dApps as custom, pluggable \gls{ran} microservices.
We focus on three key elements: (i)~the integration of dApps with \gls{ran} nodes, extended via a service-based E3 \new{interface}; (ii)~the proposed E3 procedures that enable seamless message exchange between the \ran and the dApps; and (iii)~the interactions between dApps and xApps, facilitated through our E3 interface and the O-\ran E2 interface to the \nearrt \gls{ric}.
\new{Moreover, the E3 interface requires two main components, the E3 \glsfirst{api} and E3 Agent.
The E3 \gls{api} is the software implementation of the E3 logical interface, providing the necessary functions for interaction.
Meanwhile, the E3 Agent is a entity that implements the E3 interface using the E3 \gls{api} to facilitate and manage communication.}
Further discussion on the dApps lifecycle and the role of the O1 interface is provided in Sec.~\ref{sec:life}.
Finally, Sec.~\ref{sec:prototype} describes a prototype \new{of E3 Agent} based on this architecture and the open-source \gls{oai} stack as a reference.

\begin{figure}[ht]
    \centering
    \includegraphics[width=1\linewidth]{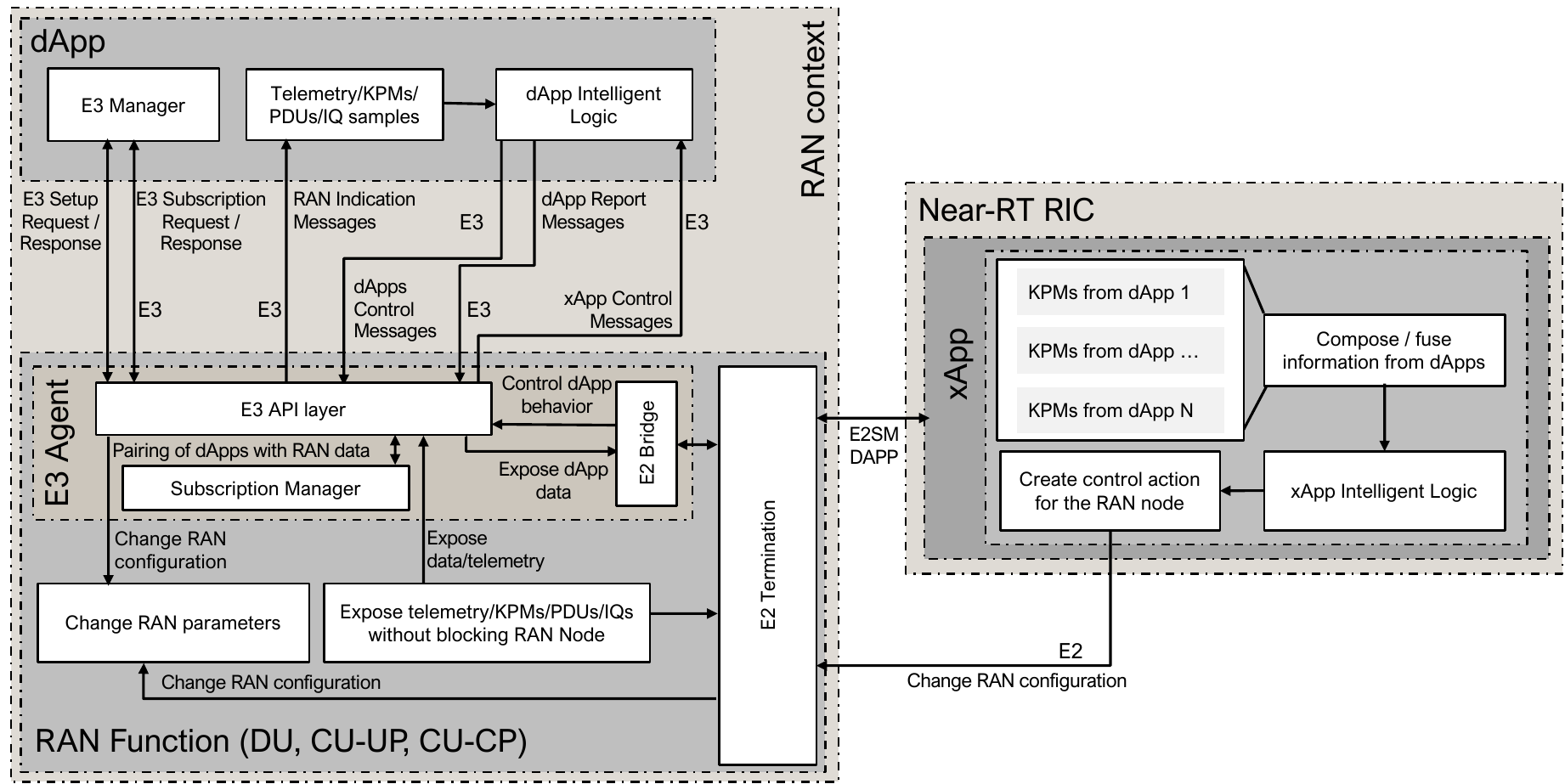}
    \caption{\oran \nearrt \gls{ric} integration with dApps and a generic \gls{ran} node.}
    \label{fig:dapp_arch}
\end{figure}

\subsection{dApps Data, Telemetry, and Control Flows}

In Fig.~\ref{fig:dapp_arch}, dApps are shown as standalone pluggable microservices with a southbound E3 interface to the \gls{ran} node (either \gls{du} and \gls{cu}) that is also bridged with the \ran E2 termination for cooperation with the xApps. The southbound interface, which we design as a service-based \gls{api}, discussed below, can be used to expose a variety of data and telemetry from the \gls{ran}. Data can be accessed through streams, by polling, and with different periodicity and granularity.

The \gls{du} \gls{ran} function can provide the dApp with access to a variety of information elements in the data plane. I/Q samples can be accessed pre- or post-equalization, depending on the dApp requirements, at different levels of granularity and periodicity. Such data can be collected from various channels such as \gls{pucch} or \gls{pusch}, and can include all subcarriers or specific subsets. 
The \glspl{bsr}, exchanged by \gls{mac} and \gls{rlc} instances, provide real-time information about buffer conditions and can be streamed at different intervals (e.g., per slot or subframe) or accessed on-demand. \gls{cqi} values are generated based on O-\gls{du} processing and can be streamed immediately or polled on demand to support dynamic resource allocation. \gls{csi} can be similarly streamed in real-time or accessed on-demand for adaptive modulation and coding decisions. Finally, uplink \gls{srs}, transmitted by the \gls{ue} and processed by the O-\gls{du}, can be made available either continuously or on request, providing insights for uplink resource configuration.

The \gls{du} and \gls{cu}-\gls{up} \gls{ran} functions can also expose transport blocks or \glspl{pdu} at the \gls{mac}, \gls{rlc}, \gls{pdcp}, and \gls{sdap} layers according to configurable policies, which may involve streaming a subset of packets at certain intervals or polling them on demand.
In the control plane, the dApps can access \gls{mac} \gls{dci} and \gls{uci} in the \gls{du}. These indications can be streamed when scheduling decisions are made, or retrieved on request to inform dApps of the current state of resource allocation. Additional information is represented by compute telemetry, i.e., information on CPU, RAM, and accelerator utilization that can be streamed at high frequency (e.g., hundreds of microseconds) to monitor system performance, or polled on demand for less frequent assessments, and fronthaul configuration, to be polled as needed to ensure that the network setup aligns with the requirements of various dApps. The \gls{cu}-\gls{cp} can expose information on slice and bearers configuration, user sessions, among others.
Finally, \glspl{kpm} from \gls{cu}-\gls{cp}, \gls{cu}-\gls{up}, and \gls{du}, which are available over E2~\cite{3gpp.28.552,3gpp.32.425,oran-wg3-e2-sm}, can also be provided to the dApps. 

In the opposite direction, dApps can provide the \gls{ran} functions with control actions, which must be applied within a short interval (e.g., 0.5\,ms) from the time a configuration is selected in the dApp. 
The areas identified for control in the \gls{ran} nodes include elements to enable the use cases in Sec.~\ref{sec:loops}. 
The following examples illustrate the capabilities of dApps but are not exhaustive and can be extended as needed. 
Within the \gls{du}, dApps can configure various aspects of the \gls{mac} scheduler, including prioritization parameters, scheduling policies, and acceleration configurations. 
They can also manage beamforming weights, codebooks, \gls{ss} blocks and bursts, and \gls{csirs}, as well as update mapping tables for \gls{sinr}-to-\gls{mcs}, apply \gls{prb} masking or nulling, and control \glspl{dci} and \glspl{uci}. 
At the \gls{cu}-\gls{up} and \gls{cu}-\gls{cp} levels, dApps enable interactions with slice and cell configurations, allowing dynamic adjustments to network slices and modifications to parameters such as transmit power and carrier frequency at the cell level.

\subsection{Service-Based dApps E3 \gls{api} Endpoint}

As depicted in Fig.~\ref{fig:dapp_arch}, dApps are deployed as services coexisting with \gls{ran} nodes. An E3 \gls{api} endpoint (or E3 agent) bridges the \gls{ran} functions with the dApps. This agent provides a standardized \gls{api} layer that supports functionalities such as dApp setup, configuration, data streaming or polling, and control. By encapsulating \gls{ran} function capabilities, the E3 agent ensures dApp independence from specific \gls{ran} implementations.
In addition to interfacing with dApps, the E3 agent coordinates with the E2 termination within the \gls{ran} node, enabling seamless communication between dApps and xApps. This facilitates the synchronization of control and adaptation processes across the \nearrt \gls{ric} and dApps at varying timescales, ensuring efficient and dynamic network management.

The architecture supports multiple dApps interacting with the same \gls{ran} function, provided there are no conflicts. Conflict resolution is addressed holistically within the network as part of the dApp lifecycle (Sec.~\ref{sec:life}). Moreover, the E3 agent, in collaboration with the O-Cloud and \gls{smo}, monitors available compute resources on the O-Cloud physical node. It ensures that \gls{ran} workloads are prioritized to meet \glspl{sla} while maintaining resource allocation for dApps.

To streamline dApp operations, the E3 agent consolidates access to \gls{ran} data, such as telemetry or data-plane units (e.g., I/Q samples).
A centralized subscription manager within the agent handles \new{the dispatching of the management messages} between the dApps and the \gls{ran}, \new{i.e., the initial pairing, the subscription request and response, and the eventual de-registration of the dApp.}
This unified access mechanism eliminates redundant protocol stack calls and ensures efficient coordination by allowing dApps to register callbacks for querying control and data functionalities.
The design simplifies interactions and maintains low latency, supporting seamless integration with \gls{ran} operations without introducing overhead.
\new{With}
The E3 \glspl{api} enable dApps to extract necessary \gls{ran} data and deliver computed control actions or inferences without disrupting \gls{ran} operations. 
The \gls{api} design ensures concurrent functionality across multiple dApps while maintaining operational integrity.
Given the co-location of \glspl{cu}/\glspl{du} and dApps on the same host, the E3 \gls{api} leverages \gls{ipc} mechanisms—such as shared memory, logical \gls{fifo} queues, function calls, or system-domain sockets—instead of network \glspl{pdu} or sockets. This approach is demonstrated in our prototype implementation (Sec.~\ref{sec:prototype}).

\subsection{dApp and RAN interactions over E3}
\label{sec:e3iface}

\begin{figure}[ht]
\centering
    \includegraphics[scale=0.5]{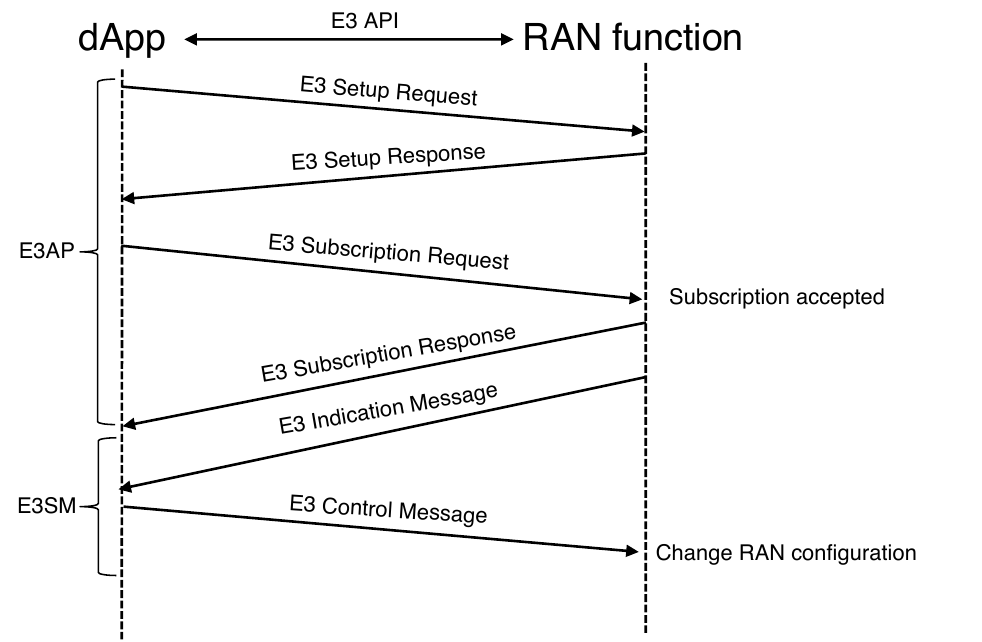}
    \caption{Message exchanges for a real time control loop between a dApp and the \ran node through the E3 interface.}
    \label{fig:e3messages}
\end{figure}

An example of the interaction flow between a dApp and the \gls{ran} E3 agent is shown in Fig.~\ref{fig:e3messages}. 
Similar to the \oran approach for the E2 interface, we introduce a procedural \gls{api}, called E3 \gls{ap} (E3\gls{ap}) consisting of two distinct logical steps to establish the association between a single dApp and a \ran node with its data.

When deployed for the first time, the dApp triggers the \textit{E3 Setup} procedure to perform the authentication and pairing with the \ran node.
Then, the \textit{E3 Subscription} procedure enables the dApp to query and subscribe to the data and control functionalities supported by the \ran function, as defined through the E3 \glspl{sm} \glspl{api}.

% As shown in the message exchange of Fig.~\ref{fig:e3messages}, 
In order to enable the exposure of data or telemetry and the control of specific \ran functionalities, the E3\gls{sm} are introduced as modular abstractions implemented through two dedicated \glspl{api}.
The first \gls{api} is for the \textit{E3 Indication Message}, to let the \gls{ran} function expose data to the dApp.
Access to this \gls{api} is coordinated by the subscription manager in the E3 agent, as discussed above.  
Similarly, the control generated by the dApp is delivered to the \gls{ran} function for processing in the stack using the \textit{E3 Control Message} \gls{api}. In Sec.~\ref{sec:use-case-1}, we will provide a practical use case where control messages from the dApp are used to avoid interfering signals via \gls{prb} nulling.

\subsection{E2SM-DAPP: a service model for managing interactions between dApps and xApps}
\label{sec:e2sm}

\begin{figure}[ht]
\centering
    \includegraphics[scale=0.5]{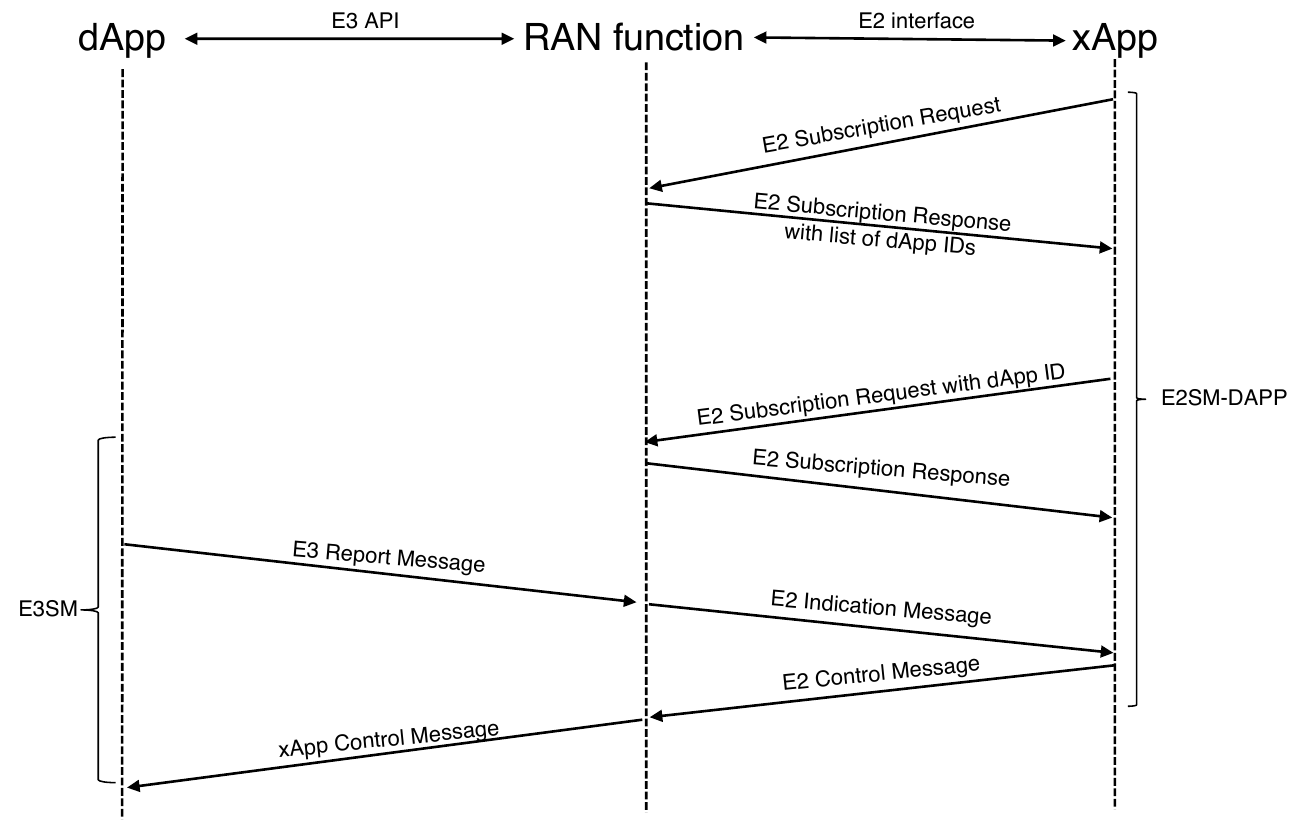}
    \caption{Interactions between a dApp, the E2 \ran node, and an xApp through the E3 interface and the E2SM-DAPP custom Service Model.}
    \label{fig:e2sm}
\end{figure}

% A dApp performs inference based on the received data. 
Apart from control actions that actively reconfigure \gls{ran} parameters based on input data, dApps can also perform forecasting and prediction tasks. The outcome of these tasks is important because it builds contextual awareness on \ran operations and conditions and, for this reason, it might be useful to make it available to higher layers of the protocol stack, e.g., xApps executing at the \nearrt \ric. This can be achieved by forwarding dApp inference outcomes to xApps in the form of enrichment information transmitted over the \gls{ran} E2 interface (e.g., similarly to how the A1 interface is used to forward useful information from the \nonrt \ric to the \nearrt \ric).
From the E2 perspective, the data exposed by dApps is managed through a novel, custom E2 \gls{sm}~\cite{oran-wg3-e2-sm}, which we call E2SM-DAPP, identified by a unique \gls{ran} function ID during the E2 Setup Request.

Fig.~\ref{fig:e2sm} summarizes the interactions between the dApps, the E2 \ran node, and the xApps.
To initiate these procedures, the xApps onboarded on the \nearrt \gls{ric} can send an E2 Subscription Request to the \gls{ran} node, which responds via an E2 Subscription Response with a message that includes the IDs of the dApps currently executing on the node.
These discovery procedures are executed continuously throughout the xApp lifecycle, enabling the discovery of new dApps as they become available.

% xApps willing to receive inference data from the dApp, can then send an additional E2 Subscription Request with the same dApp ID and a payload specifying the data that they want to collect from the dApp, or which policy they want to control\hl{which policy? do you mean that the xApp can generate a high-level policy that changes the way the dApp behaves? If yes, this needs to be clear. You can say that xApps can also configure dApp decision making by specifying high-level parameters and policies that the dApp uses to take decisions (e.g., priority weights in the case of a dApp controlling scheduling of UEs)}.
xApps that are willing to receive inference data (e.g., spectrum sensing outcomes, real-time channel estimations, scheduling information) from the dApp can send an additional E2 Subscription Request with the same dApp ID and a payload specifying the data they want to collect from the dApp. 

Once this pairing is established, the dApp periodically collects real-time data from the \ran using the real-time control loop procedures described in Sec.~\ref{sec:e3iface}, performs near-real-time inference (e.g., spectrum classification), and transmits these inferences to the xApp using the \textit{E3 Report} \gls{api}, which connects the dApp to xApps leveraging an E2 Indication Message via the \ran E2 interface.

It is also worth mentioning that xApps can configure how dApps take decisions and perform inference by specifying high-level parameters and policies. 
These policies influence how the dApp operates and can be used by xApps to influence decisions. For example, in the case of a dApp performing scheduling operations, an xApp can specify which scheduling profile to use (e.g., round-robin, proportional fairness) as well as defining priority weights for scheduling \glspl{ue} and prioritizing certain types of traffic. To enable this, the xApp sends an E2 Control Message to the \ran node, which is then relayed over the E3 interface using the \textit{xApp Control Message} to the target dApp.

\section{dApp Lifecycle and Interaction with \nearrt RIC, Non-RT RIC, and SMO}
\label{sec:life}
% \hl{Niloofar}

%In this section, we discuss how the \gls{lcm} of dApps can be integrated in \gls{lcm} practices defined within the \oran architecture and processes. We follow and extend the specifications outlined in ~\cite{oran-wg10-OAM, oran-wg6-ORCHCL}, documents that describe the \oran applications \gls{lcm} and cloudification and orchestration use cases for deploying \oran virtualized \gls{ran}.
In this section, we discuss how the \gls{lcm} of dApps can be integrated into \gls{lcm} practices defined within the \oran architecture and processes. We follow and extend the specifications outlined in~\cite{oran-wg10-OAM, oran-wg6-ORCHCL}, which describe the standard procedures for the \gls{lcm} of \oran applications, cloudification, and orchestration use cases for deploying virtualized \oran solutions. 
%
% The \gls{lcm} of dApps within the \oran ALLIANCE follows the specifications outlined in the document \oran Cloudification and Orchestration Use Cases and Requirements for \oran Virtualized \gls{ran}. 
The \gls{lcm} process follows a structured 7-stage model for the \gls{sdlc} to ensure that applications are developed, onboarded, and operated in a standardized way. It begins with \textit{Need} to identify requirements, followed by \textit{Ideation} to brainstorm potential solutions, \textit{Analysis} to assess feasibility, \textit{Develop} to build the application, \textit{Deliver} to deploy the solution, \textit{Validate} to test for quality, and finally \textit{Operate} to maintain the application in production.

\begin{figure}[ht]
    \centering
    \includegraphics[width=\linewidth]{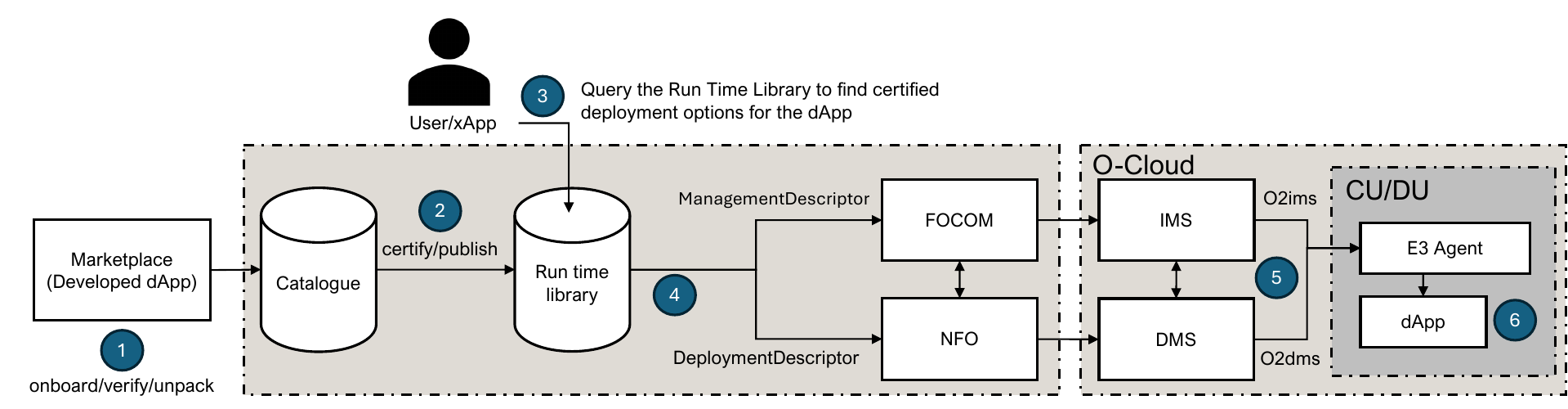}
    \caption{dApps \gls{lcm} diagram.}
    \label{fig:dapp_lcm}
\end{figure}

This structured \gls{lcm} approach is divided into three main phases: (i)~Development, in which applications are designed and built; (ii)~Onboarding, in which applications are integrated into the target environment; and (iii)~Operations, in which applications are managed and maintained in production~\cite{oran-wg10-OAM}. 
Fig.~\ref{fig:dapp_lcm} provides a high-level representation of dApp \gls{lcm}, with particular focus on the onboarding and operations phases.
While the development process is common to all types of applications (i.e., xApps, rApps, and dApps), Fig.~\ref{fig:dapp_lcm} focuses specifically on unique aspects related to onboarding and operation of dApps. 
\new{While the \gls{smo} is responsible for deploying the dApp, once deployed, the dApp can operate independently or be managed by other entities, such as an xApp, through the E2SM-DAPP.}
In the following, we provide a detailed overview of how these phases can be extended and adapted to dApp \gls{lcm}.

\subsection{Development}

The development phase of the dApp \gls{lcm} aligns with the Solution App Lifecycle and Solution AppPackage Lifecycle outlined in the \gls{lcm} model in~\cite{oran-wg10-OAM}. The Solution App Lifecycle covers the end-to-end process of designing, developing, deploying, and managing the dApp, ensuring its smooth operation across its entire lifecycle. The Solution AppPackage Lifecycle focuses specifically on the packaging, onboarding, and management of the application package, including the creation of deployment and management descriptors to standardize the process. The process begins with defining the use case, requirements, and features, followed by the integration of components such as intelligent logic, telemetry, and \glspl{kpi} of the dApp. The dApp is then containerized using tools like Docker for efficient deployment in cloud environments such as Kubernetes or OpenShift. Once containerized, the Solution AppPackage is created to ensure proper onboarding, containing key descriptors like the DeploymentDescriptor and ManagementDescriptor, which standardize deployment and management processes for seamless orchestration.

To ensure secure onboarding, digital signatures and checksums are added to verify the package source, preventing unauthorized modifications and ensuring the application integrity as it transitions through different environments.

\subsection{Onboarding}

% In the onboarding phase, the dApp AppPackage, generated during the development phase, is onboarded transferring it from the marketplace where it was originally stored after development. The dApp package undergoes verification and validation to ensure it is authorized and secure. Once validated, the AppPackage is unpacked, and its components are stored in a catalog within the \gls{smo}, as illustrated in Step 1 of Fig.~\ref{fig:dapp_lcm}. Each recommended configuration of the dApp is then certified before being published to a runtime library (Step 2).
% The initiation of the dApp onboarding process can be triggered directly by an operator, who queries the Run-Time Library to find certified deployment options for the dApp
The onboarding phase begins with the transfer of the dApp AppPackage, generated during the development phase, from the marketplace where it was stored. This package undergoes verification and validation to ensure it is authorized and secure. Once validated, the AppPackage is unpacked, and its components are stored in a catalog within the \gls{smo} (Step 1, Fig.~\ref{fig:dapp_lcm}). Following this, each recommended configuration of the dApp is certified and published to a runtime library (Step 2).

The initiation of the onboarding process can typically be triggered either by the \gls{smo} or directly by an operator, who queries the runtime library to locate certified deployment options for the dApp. In certain scenarios, this decision may also be delegated to xApps or rApps if they are equipped with the necessary intelligence and reliability to orchestrate and control dApp execution. Indeed, it can be advantageous for an xApp to be able to initiate the deployment of dApps. This allows the xApp to use the available dApp for tasks requiring real-time control. The process flow involves the xApp querying the available dApps in the Run-Time Library through the O1 interface, and requesting the dApp deployment via the O1 interface (Step 3). 

However, in Fig.~\ref{fig:dapp_lcm}, the process for xApp-initiated onboarding is not explicitly shown as this procedure requires ad-hoc service models, procedures and functionalities that are not yet available in the \oran specifications, and we hope to cover in future research.

\subsection{Deployment}

In contrast to the deployment of xApps and rApps, hosted in the \nearrt and \nonrt \glspl{ric}, dApps require a different deployment procedure, as they are directly hosted in the \gls{du}/\gls{cu}. dApps are deployed as \gls{cnf}, adhering to the standard \gls{nf} deployment procedures set out by the \oran ALLIANCE~\cite{oran-wg6-ORCHCL}. 

Using the \gls{smo} framework and the O-Cloud, the dApp is managed through the O2 interface, which facilitates communication between the \gls{smo} and the \gls{dms} of the CU/DU. This direct deployment model allows the dApp to utilize the real-time processing capabilities of the CU/DU while also leveraging the flexibility and scalability of the O-Cloud infrastructure. However, this deployment diverges significantly from that of xApps and rApps, as it is executed outside of the \gls{ric} environments, focusing instead on efficient resource allocation and real-time operations directly at the CU/DU level. This approach integrates dApps seamlessly into the CU/DU environment while adhering to the cloud-native principles and ensuring optimal network function performance. The deployment steps of dApps in the \gls{lcm} workflow are shown in Steps 4, 5, and 6 in Fig.~\ref{fig:dapp_lcm}. 

The information flow for deploying the dApp follows a structured sequence, beginning with a service request initiated by the Network Function Install Project Manager to the \gls{smo} for deploying a new dApp instance on the CU/DU. The \gls{smo} processes this request by decomposing it, identifying the required dApps and their deployment order, and determining the deployment parameters based on policies or explicit input. The \gls{smo} retrieves the CloudNativeDescriptor for the dApp from the runtime library and directs the O-Cloud Deployment Management Service (DMS) to create the dApp deployment. 

The DMS allocates the necessary compute, storage, and network resources on the CU/DU, deploys the dApp container(s), and notifies the \gls{smo} upon successful instantiation, providing a Deployment ID. The \gls{smo} then updates its inventory with the deployment status, and the dApp instance reads its configuration and begins operation. Following deployment, the \gls{smo} continuously monitors the dApp’s health, performance, and connectivity, managing its lifecycle, scaling, and updates as required. Finally, the \gls{smo} informs the Network Function Install Project Manager of the overall success or failure of the deployment request.
For a detailed breakdown of these steps, please refer to Table~\ref{tab:lcm} in~\ref{appendix:dapp-deployment}.

Finally, we also consider the case where xApps can also request the \gls{smo} to deploy a new dApp, and connect to it, whenever there is a need to perform real-time control. Deployment is performed following the procedures described above.

\section{A Reference Open-Source dApp Framework}
\label{sec:prototype}

In this section, we present a reference implementation and prototype of the proposed dApp architecture (Sec.~\ref{sec:arch}), based on \gls{oai} and a Python framework. We also present an exhaustive benchmarking analysis with results that confirm \ran 
control loop latencies well below 1\,ms (i.e., below the target threshold of 10\,ms). The reference implementation supports real-time control loops. The open-source library can thus be used for real-time Open \ran control prototypes. The use-cases developed as a proof of concept for our framework and their evaluation are discussed in in Sec.~\ref{sec:use-cases}.
% as a reference.

\subsection{dApp Framework}
\label{sec:dappframe}

\begin{sidewaysfigure}
    \centering
    \includegraphics[width=\textwidth]{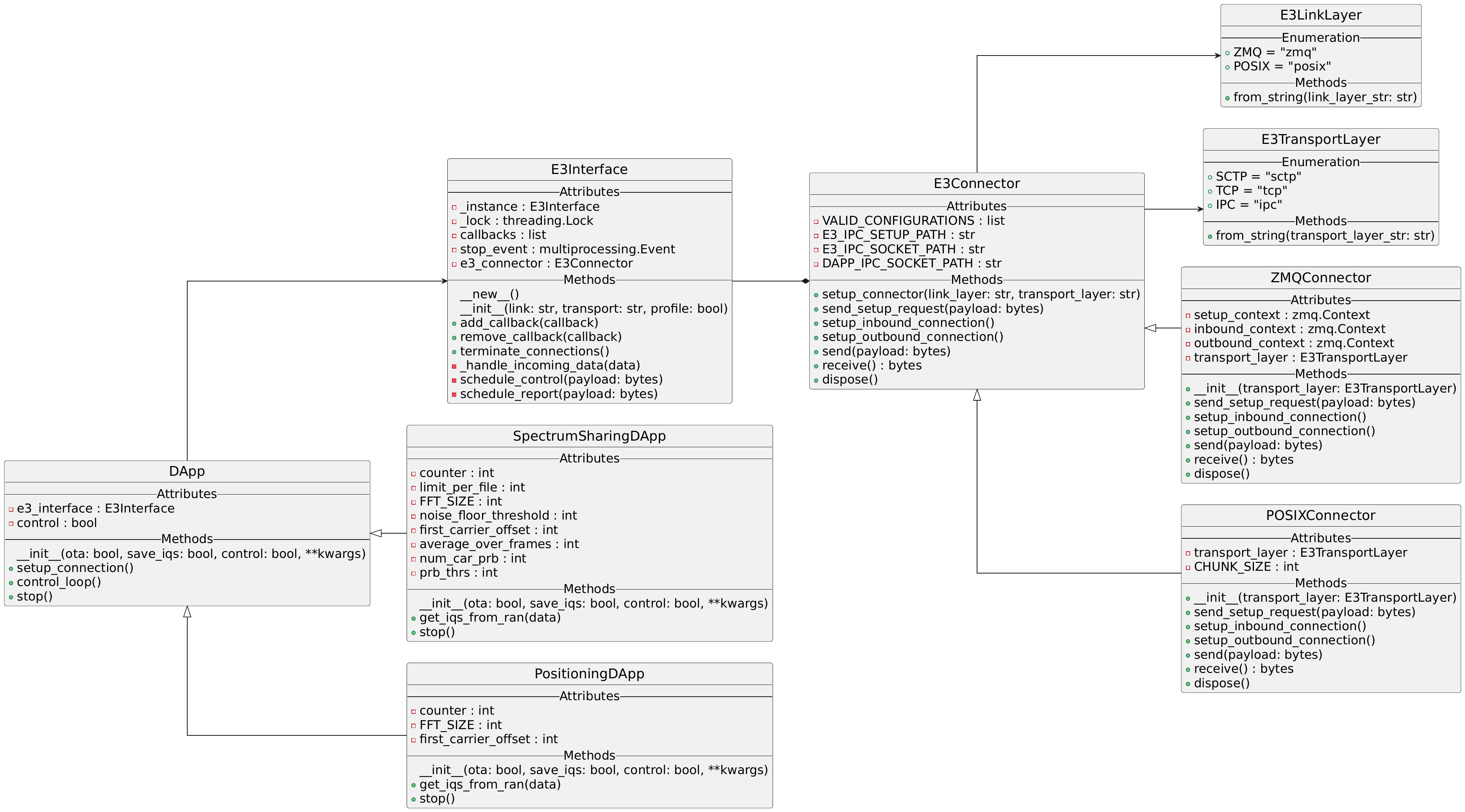}
    \caption{\gls{uml} diagram of the dApp framework with its components and the two dApps presented in this work used as example.}
    \label{fig:uml}
\end{sidewaysfigure}

We define a reference dApp framework written in Python, implementing the E3\gls{ap} as described in previous sections. 
Fig.~\ref{fig:uml} illustrates the \gls{uml} diagram of our framework. 
We design three main classes composing the dApp framework to implement diverse functions: the \texttt{DApp}, the \texttt{E3Interface}, and the \texttt{E3Connector} classes. 

{\bf The DApp Class --} First, we define the the \texttt{DApp} abstract class for the management of the E3\gls{sm} operations. For each use case, this class will be extended to include functionalities, parameters and operations specific to the use case.
This class wraps all the E3\gls{sm} functionalities into a single entity to simplify and streamline the generation of new children classes and enable the extension to new use cases.
The \texttt{SpectrumSharingDApp} and the \texttt{PositiningDApp} reported in Fig.~\ref{fig:uml} are two examples of children classes implementing the operations reported in the the spectrum sharing use case of Sec.~\ref{sec:use-case-1} and the positioning use case of Sec.~\ref{sec:use-case-2}.
The abstract \texttt{DApp} class uses the \texttt{E3Interface} class for the communication between with the \gls{ran} node.

The abstract \texttt{DApp} class leverages the \texttt{E3Interface} class to facilitate communication with the \gls{ran} node.
The E3\gls{ap} operations and interactions between the \gls{ran} and dApps utilize a publish-and-subscribe mechanism, where dApps register callbacks to access \gls{ran} data through specific \gls{ran} function IDs.
These data are published via the \texttt{E3Interface}, which is a private variable of the abstract class, as illustrated in Fig.~\ref{fig:uml}.
The \texttt{setup\_connection()} class method initializes the E3 interface and the dApp callback, while the \texttt{control\_loop()} method is designed to implement the data polling and periodically trigger the core logic of the dApp.

{\bf The E3Interface Class --} The \texttt{E3Interface} class provides standardized \glspl{api} for accessing the E3 interface. 
It manages the E3\gls{ap} and facilitates the dispatch of data from the \gls{ran} node to dApps, enabling both analysis and control operations via function calls. 
The class implements a singleton pattern, ensuring that all dApps deployed on the same \gls{ran} node share the same E3 interface.

The \texttt{DApp} class interacts with the \texttt{E3Interface} by invoking methods such as \texttt{add\_callback()} and \texttt{remove\_callback()} to manage subscriptions to \gls{ran} data. 
Additionally, it uses \texttt{schedule\_control()} to schedule the delivery of E3 Control Actions to the \ran and \texttt{schedule\_report()} to send E3 Report Messages to xApps. 
Once a dApp has initiated its control loop, the \texttt{E3Interface} exposes \ran data to the dApps through the \texttt{handle\_incoming\_data()} method.

\begin{figure}[ht]
\centering
    \includegraphics[width=1\linewidth]{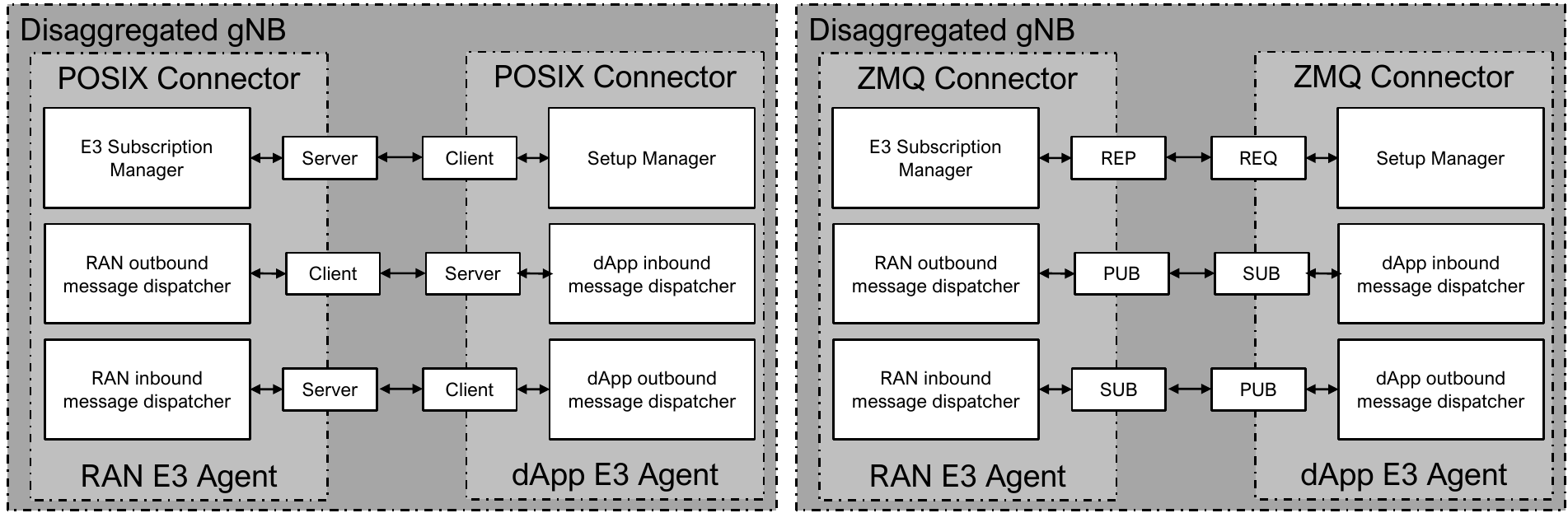}
    \caption{Design of the E3Connector sockets.}
    \label{fig:e3connector}
\end{figure}

{\bf The E3Connector Class --} Ensuring fast and reliable connectivity between dApps, the RAN and xApps is crucial to enable dApp operations and real-time monitoring and control of the network. For this reason, to understand what are the trade-offs and the best protocols to achieve real-time control loops, we have implemented and compared several connectivity solutions and protocols. These efforts have been synthesized in the \texttt{E3Connector} class, which is an abstract class that the \texttt{E3Interface} uses to create the socket connections between the dApps and the \ran node.
The \texttt{E3Connector} exposes a set of methods that are used to perform the initial pairing between the dApp and the \ran node, receive the messages from the \ran node and send the messages to it.
As shown in Fig.~\ref{fig:e3connector}, which represents the different logical pairings between the configurations of the connectors according to the protocols used, the \texttt{E3Connector} creates and manages three different sockets: the first one is for the initial E3 Setup Request-Response and the E3 Subscription Request-Response exchanges; the second one for receiving the inbound messages from the \ran unit, i.e., the E3 Indication Message and the xApp Control Message, and the last one for delivering to the \ran node the outbound messages of the dApps, i.e., the dApp Control Messages and the dApp Report Messages.

In this work, we have created two mutual children classes of the \texttt{E3Connector}, the \texttt{ZMQConnector} and the \texttt{POSIXConnector}, each one implementing a different data link layer and several transport layers.
The first one is the \texttt{POSIXConnector} implements the message exchange using the classic POSIX system functions based on the Linux kernel socket implementation.
Such functions expose low-level \glspl{api} for \gls{ipc} like shared memory, pipes, message queues, and sockets and they require detailed handling for synchronization, data serialization, and transport, focusing on one system or tightly coupled systems.
All the sockets implemented by the \texttt{POSIXConnector} follow a client-server pattern, having the receiving entity set as a server and the delivering entity acting as a client.

The \texttt{ZMQConnector} class uses ZeroMQ as the link layer, an high-performance asynchronous messaging library, aimed at use in distributed or concurrent applications~\cite{hintjens2013zeromq}.
ZeroMQ abstracts the complexity of operations such as connection management and data serialization via simple \glspl{api} calls, and it is designed for both intra- and inter-machine communication with flexible built-in patterns, as discusses below.
As illustrated in Fig.~\ref{fig:e3connector}, our approach leverages ZeroMQ. Specifically, the initial E3 setup employs a request-reply (REQ-REP) pattern—where the dApp issues the request (REQ) and the \ran side provides the reply (REP)—since this procedure occurs only once during the dApp’s lifecycle. In contrast, subsequent outbound and inbound data exchanges adopt a publish-subscribe (PUB-SUB) pattern, with the data-generating entity acting as the publisher (PUB) and the data-processing entity as the subscriber (SUB). This choice reflects the need to support continuous data transfer between the \ran and the dApp, in contrast to the one-time setup operation.

Both classes implement three transport-layer protocols—\gls{tcp}, \gls{ipc} (using Unix Domain sockets), and \gls{sctp}—to facilitate communication between local and remote hosts.  While the implementation can support other transport protocols, in our design we focus on those that offer reliable message delivery (even if this introduces overhead) and excludes those, e.g., \gls{udp}, that do not guarantee data delivery and are not reliable enough for \ran operations.
Furthermore, to maintain consistency, the \texttt{ZMQConnector} class includes \gls{sctp} support, despite the ZeroMQ link layer not yet offering full \gls{sctp} compatibility~\cite{hintjens2013zeromq,zeromq-sctp}.
It is also worth mentioning that we decided to maintain the possibility to have a connection for the E3 to a remote virtual host, not co-located with the \ran node, since in the \oran vision the disaggregation of the network functions may also happen across diverse physical nodes in the network. 
% As such, the \texttt{E3Connector} supports by design this case and others such as \gls{iab}.

Data received through the \texttt{E3Connector} socket, regardless of the underlying link and transport layers, is processed by a callback mechanism operated by the subscription manager. This mechanism associates newly available data with the registered dApps that require it, thus enabling real-time data handling and avoiding data duplication.
% In such way, the dApp implementation builds on the \texttt{E3Interface} for performing real-time data inference and control. 
For the \glspl{pdu} used for establishing communications between the \ran node and dApps, we employ \gls{asn1} definitions for message formatting in the E3\gls{ap} and E3\gls{sm}, adhering to established standard \gls{3gpp} and \oran message exchange protocols. \gls{asn1} enables efficient, rapid, and compact serialization of data.
Each class extending our \texttt{DApp} abstract class should implement its own \gls{sm} and intelligent logic according to its use case, as shown in Fig.~\ref{fig:uml}.

In this work, we propose two different use cases leveraging dApps based on this framework with further implementation details discussed in Sec.~\ref{sec:use-cases}.

\subsection{OpenAirInterface and T-tracer}
\label{sec:oaiframe}

\gls{oai} is an open-source project that implements a \gls{3gpp}-compliant full \gls{5g} \gls{nr} stack  on general-purpose computing hardware and off-the-shelf \glspl{sdr}~\cite{kaltenberger2024driving,KALTENBERGER2020107284}
It supports both \gls{cu}-\gls{du} split and monolithic deployments, enabling flexibility in testing and development of \gls{ran} functions. 
Additionally, \gls{oai} provides implementations for key \gls{ran} components, including \gls{gnb} and \gls{ue}, offering a robust platform for validating \gls{5g} network functionalities and integrating advanced features such as slicing and \gls{ai}/\gls{ml}-based optimizations.

The project
% is well-maintained to this day and
implements Open \gls{ran}-compliant functionalities, including code functions to connect with the \nearrt \gls{ric} (both the \gls{osc}~\cite{bimo_osc_2022} version and custom versions such as FlexRIC~\cite{10.1145/3485983.3494870}) and the \nonrt \gls{ric}.

We have extended the original \gls{oai} codebase by introducing a new E3 Agent module that facilitates exchanges based on the architecture proposed in this paper, with a particular attention to the message exchange between the dApp and the xApps over the E3\gls{ap}, and the \gls{ran} control functionalities employed by the proposed use cases. 

It is worth mentioning that \gls{oai} integrates T-tracer~\cite{mundlamuri20245g}, a programmable toll that provides an external connection via \gls{tcp} sockets, enabling developers to extract and collect metrics during \gls{ran} operations for debugging and analysis purposes. 
Our module integrates with the T-tracer tool, enabling real-time data sensing and extraction, and making this data accessible through the E3 interface. 
We extended the T-tracer socket utilities to support \gls{ipc} via Unix domain sockets, enhancing the communication and reducing the overhead generated in the data exchange and, consequently, the latency. 
Moreover, we developed a C-language implementation of the \texttt{E3Connector} to integrate it into the \gls{oai} codebase through a merge request, avoiding external dependencies between the dApp framework and cellular stack. This component works in the very same way as its Python counterpart, enabling different link- and transport-layer tests over different virtual hosts by design.
We will propose the integration of this connector in the \gls{oai} main codebase upon acceptance of this manuscript.
% with no software change.
Finally, we extended the \gls{oai} codebase to implement the data extraction and the \ran control of the use cases discussed in Sec.~\ref{sec:use-cases}, supporting real-time control loops. 

\subsection{Evaluating real-time capabilities of dApps}
\label{sec:loop_benchmark}

We ran more than two hundreds experiments on two different testbeds to assess and evaluate the capabilities of our framework executing the \texttt{SpectrumSharingDapp}. A detailed discussion of the use case is  provided in Sec.~\ref{sec:use-case-1}. In the following paragraphs we focus specifically on the benchmarking of the framework in terms of real-time control capabilities. 

The first test bench for this evaluation is the Colosseum testbed, a wireless network emulator~\cite{polese2024colosseum} consisting of \gls{sdr}-equipped servers and a channel emulator capable of reproducing real-world wireless channel effects, such as path loss and fading. 
The second is Arena~\cite{BERTIZZOLO2020107436}, a publicly available over-the-air indoor testbed deployed in an laboratory environment. 
Both systems used the same version of \gls{oai} and our dApp prototype, running on Dell PowerEdge R730 servers and \gls{usrp} x310 \glspl{sdr}.
On Arena, the \gls{ue} is a One Plus Nord 5G AC2003, while on Colosseum we use the \gls{oai} softUE.
We use the \gls{oai} configuration for the \gls{fr1} band n78, with a center frequency of 3.6192\,GHz and a bandwidth of 40\,MHz. 
%
% The key distinction is that Colosseum operates through an emulated channel, while Arena is an over-the-air testbed.

Each experiment has a duration of six minutes.
However, for the scope of the measurements reported, we only consider the five-minute window where \gls{ue} downlink transmissions are active (i.e., we exclude the network bootstrap, \gls{ue} attachment and setup phases).
We present the results related to the performance of the real-time control loop.
\begin{table}[ht]
    \centering
    \begin{tabular}{|c|c|c|}
    \hline
     \textbf{Protocol} & \textbf{Overhead percentage (no ASN)} & \textbf{Overhead percentage (with ASN)}  \\ \hline
        TCP & 20\%  & 20.05\% \\ \hline
        SCTP & 42\% &  42.05\% \\ \hline
        \gls{ipc} & 0\% & 0.05\% \\ \hline
    \end{tabular}
    \caption{Overhead introduced in the E3AP message exchange using the \texttt{E3Connector}.}
    \label{tab:control_loop}
\end{table}
%
% As stated in Sec.~\ref{sec:prototype}, during the design phase,
We evaluated different options for minimizing overhead, focusing on the link layer and the transport layer. Specifically, we considered the two
\texttt{E3Connector} implementation options based on raw POSIX functions and ZeroMQ.
Moreover, we have implemented and evaluated three different options for the transport layer built on top of these two links, i.e., the \gls{tcp} protocol, the \gls{sctp} protocol and the \gls{ipc} based on Unix Domain sockets.

Table~\ref{tab:control_loop} presents a summary of the overheads introduced by the different transport layers protocols, i.e., the number of bytes introduced by the protocol stack to guarantee message delivery.
% The reported overheads do not account for potential packet losses and are fixed, being directly proportional to the number of \glspl{pdu} exchanged through the E3 interface. 
% To properly evaluate the impact that overhead has on latency, we executed experiments with different \gls{pdu} configurations where we vary the size of both control (e.g., the size of control actions sent over E3) and indication messages as reported in Fig.~\ref{fig:loop_stats}.
% Since the duration of the experiments is also fixed, these overheads remain constant.
The first protocol we consider is \gls{tcp}, which is commonly used for inter-process communication of application that can scale over different virtual hosts.
The additional overhead caused by \gls{tcp} is 20\%, making it a good candidate in the case where the dApp is connected across virtual hosts over the same network.
As a second candidate, we evaluated the \gls{sctp} protocol since it is an affirmed transport-layer protocol for cellular networks.
Results show that \gls{sctp} introduces significant overhead, which reduces goodput, increases latency, and can potentially slow down the control loop process and result in late control action enforcement. 
While this overhead is essential for other message exchange procedures in \oran, such as E2\gls{ap}, dApps can instead leverage \gls{ipc}. 
This approach eliminates overhead entirely, as \gls{ipc} allows direct communication between processes on the same host. 
Unlike network-based protocols, \gls{ipc} avoids the need for data encapsulation and transmission across multiple network hosts, significantly reducing latency and processing requirements.
Indeed, \gls{sctp} brings many benefits such as multi-homing support, path \gls{mtu} discovery, redundant transmission for reliability.
However, these come with additional overhead, e.g., due to encapsulation and decapsulation of link and transport layers, that can slow down the flow of the dApp, which make \gls{ipc} a better candidate to deliver real-time control over the same host.

Finally, we report the overhead generated by the \gls{asn1} implementation of the E3\gls{ap}-\gls{sm} \glspl{pdu} application layer.
In the design of such definitions, we opted for a minimalist approach that ensures the correct implementation of the architectural framework by providing only the necessary \glspl{pdu} (i.e., Setup, Subscription, Indication, and Control) to enable a closed control loop, while using an architecture that can be extended with more \glspl{pdu} in the future if needed.
The \gls{asn1} implementation offers a structured messaging infrastructure, but \gls{asn1} structures and its padding generates overhead.
This overhead is the necessary trade-off for representing the advanced logical interactions between the dApps and the \gls{ran} through the E3\glspl{sm}.
To reduce overhead, in our implementation we use the \gls{per}, which is a more compact representation of \gls{asn1} that minimizes the overhead by omitting extra information such as tags and lengths of packet fields.
For instance, if fields have a limited number of possible values, the encoding will use just enough bits to represent those values, making the encoded data smaller compared to other encoding rules.
In our experiments, the \gls{asn1} \glspl{pdu} generate a constant 0.05\% overhead across all protocols, which adds to the overhead generated by each protocol.

\begin{figure}[!ht]
\centering
\begin{subfigure}[t]{0.5\linewidth}
    \scalebox{0.25}{\input{images/loop_stats/1536.pgf}}
    \caption{384 I/Q samples - Indication size = 1536 bytes}
     \label{fig:realtime:1536}
\end{subfigure}\hfill% equal to outside spacing
\begin{subfigure}[t]{0.5\linewidth}
    \scalebox{0.25}{\input{images/loop_stats/3072.pgf}}
    \caption{768 I/Q samples - Indication size = 3072 bytes}
         \label{fig:realtime:3072}
\end{subfigure}
\begin{subfigure}[t]{0.5\linewidth}
    \scalebox{0.25}{\input{images/loop_stats/6144.pgf}}
    \caption{1536 I/Q samples - Indication size = 6144 bytes}
         \label{fig:realtime:6144}
 \end{subfigure}\hfill% equal to outside spacing
\begin{subfigure}[t]{0.5\linewidth}
   \scalebox{0.25}{\input{images/loop_stats/8192.pgf}}
    \caption{2048 I/Q samples - Indication size = 8192 bytes}
     \label{fig:realtime:8192}
\end{subfigure} 
\caption{Latency measurements for the real-time control loops using dApps with E3AP over ZeroMQ and \gls{ipc}.}
\label{fig:loop_stats}
\end{figure}
%% Add paragraph to explain the operations
% Fig.~\ref{fig:loop_stats} reports the measured latency experienced when generating control messages from the dApp to the \ran of \gls{oai}. 
% To evaluate how the different dApp operations affect latency, in the figure we dissect the real-time control loop into four different operations and we provide results for each one individually.
% Upon the generation of the \ran measurements, the dApp shall collect such data from the \ran via the T-tracer using the E3 Indication Message, then the dApp should process the data through an heuristic or a data driven approach and then it should create the control action and the E3 Control Message  and finally such message will be delivered to the \ran function.
% such as \hl{explain here what each element is. You explain it later on, but it needs to be the first thing you explain} data retrieval , \hl{etc} .

Fig.~\ref{fig:loop_stats} reports latency measurements for all operations required to perform dApp-based control using our prototype. 
Specifically, we measure the latency required to perform the four fundamental operations of dApp control: \textit{Collect Data}, \textit{Process Data}, \textit{Create Control}, and \textit{Deliver Control}. We also report the \textit{Cumulative} latency, defined as the sum of all the previous operations.
Initially, upon the generation of the \ran measurements, the dApp must \textit{Collect Data} from the \ran using the T-tracer and the \textit{E3 Indication Message}.
Subsequently, it moves to \textit{Process Data}, where the measurements are analyzed through either a heuristic or data-driven approach. 
Following this, the dApp proceeds to \textit{Create Control}, which involves generating the control action and formulating the \textit{E3 Control Message}.
Finally, in the \textit{Deliver Control} stage, this message is transmitted back to the \ran function. 

To evaluate our prototype in a practical use case that relies on real-time operations and inference, we provide results obtained by executing the spectrum sharing dApp (described in Sec.~\ref{sec:use-case-1}). As we will discuss later,  the input data to be retrieved consists of I/Q samples, and the output of the dApp corresponds to the list of \glspl{prb} that the \gls{mac} scheduler should not allocate due to the presence of external interference (e.g., incumbents or jammers). We also consider the implementation that provided the best performance which is the case of an \texttt{ZMQConnector}-based E3 interface serving data through \gls{ipc} and \gls{asn1}, and provide results for varying dApp input and output sizes.
Specifically, we considered 16\,different configurations by varying two parameters: (i) the size of the payload in the \textit{E3 Indication Messages} (i.e., input of the dApp); and (ii) the size of the \textit{E3 Control Messages} (i.e., output of the dApp). 
For the Indication Messages, each I/Q sample is encoded using 4 bytes, and we evaluated four payload sizes of 1536, 3072, 6144, and 8192 bytes, which correspond to 384, 768, 1536, and 2048 I/Q samples, respectively.
For the Control Messages, we defined payload lengths in bytes based on the number of \glspl{prb} blocked by the dApp of the aforementioned use case: 0 (no \glspl{prb}), 4 \glspl{prb} (16\,bytes and about 1\,MHz of blocked bandwidth), and 8 \glspl{prb} (32\,bytes and about 2\,MHz). It is worth mentioning that the list of \glspl{prb} to be blocked is of variable size and depends on the resolution of the Indication Message (i.e., how many I/Q samples in the frequency domain are fed to the dApp). For this reason, we also consider the maximum number of \glspl{prb} that can be blocked based on the amount of I/Qs being included in the Indication message size. Specifically, for Indication Message size of 384, 768, 1536, and 2048 we consider Control Messages of maximum size 128, 256, 208, and 380\,bytes, respectively. 
% More details on this are provided in Sec.~\ref{sec:use-case-1}.

% In the case of Fig.~\ref{fig:realtime:6144}, which is also the case we used for the performance evaluation of the \gls{gnb}-\gls{ue} communication in Sec~\ref{sec:use-case-1}, the maximum value is set to a lower value to avoid blacklisting the \glspl{prb} where the carrier \gls{bwp} is scheduled.
% This is a workaround because otherwise the \gls{ue} and \gls{gnb} would not be able to communicate anymore.
% A better solution would be to move the \gls{bwp} as needed, which is beyond the scope of this paper.
% Please note that a control message with a list equal to zero is used to signal to the \gls{ran} that the incumbent is not present anymore in the spectrum.
% %
% Moreover, the byte sizes reported represent the actual logical information exchanged between the \gls{ran} and the dApp, i.e., \textit{prior} to the encapsulation process of the \gls{asn1} previously described. 
% Consequently, the final amount of bytes exchanged over the E3 interface is larger than the values reported here since it includes the overheads summarized in Tab.~\ref{tab:control_loop}.

Results in Fig.~\ref{fig:loop_stats} demonstrate that our dApp prototype achieves real-time control loops with an average aggregated latency of 400\,$\mu$s, consistently remaining below 450\,$\mu$s.
As expected, the usage of \gls{ipc} through ZeroMQ substantially reduces the impact of the socket operations on the total control loop time, making the operation of \textit{Data Collection} the indication message and \textit{Control Delivery} the fastest operations in the loop.
The major contribution to the elapsing time of the loop is represented by the \textit{Elaboration} of the Indication Message, i.e., the \gls{asn1} decode and the intelligent control logic implemented by the dApp to analyze the data.
This behavior is also highlighted in the case of Control Messages with size 64 and 95\,bytes, shown in Figs.~\ref{fig:realtime:3072} and~\ref{fig:realtime:8192}, respectively, where the presence of padding in the E3\gls{ap} \gls{pdu} causes additional memory allocation, thus increasing the elaboration time.
The average total elapsed time for each of the 16~cases is consistently below the 1\,ms threshold as shown in Fig.~\ref{fig:loop_stats}, which is defined as the limit for real-time use cases involving the manipulation of I/Q samples (see Fig.~\ref{fig:loops}, Sec.~\ref{sec:loops}).

% In the following sections, we will explore the new capabilities, software extensions, and control mechanisms that dApps practice to the \ran.

\section{Empowering \ran control through dApps: use cases}
\label{sec:use-cases}

After demonstrating that dApps can indeed enable real-time inference in \oran systems, in this section, we showcase two relevant \ran control use cases that benefit from dApps. These examples illustrate how the data extraction and dApp-driven control capabilities provided by our approach can enhance network performance and make network management more agile.

\subsection{Use Case: Spectrum Sharing}
\label{sec:use-case-1}

\begin{figure*}[ht]
    \centering
    \includegraphics[width=1\linewidth]{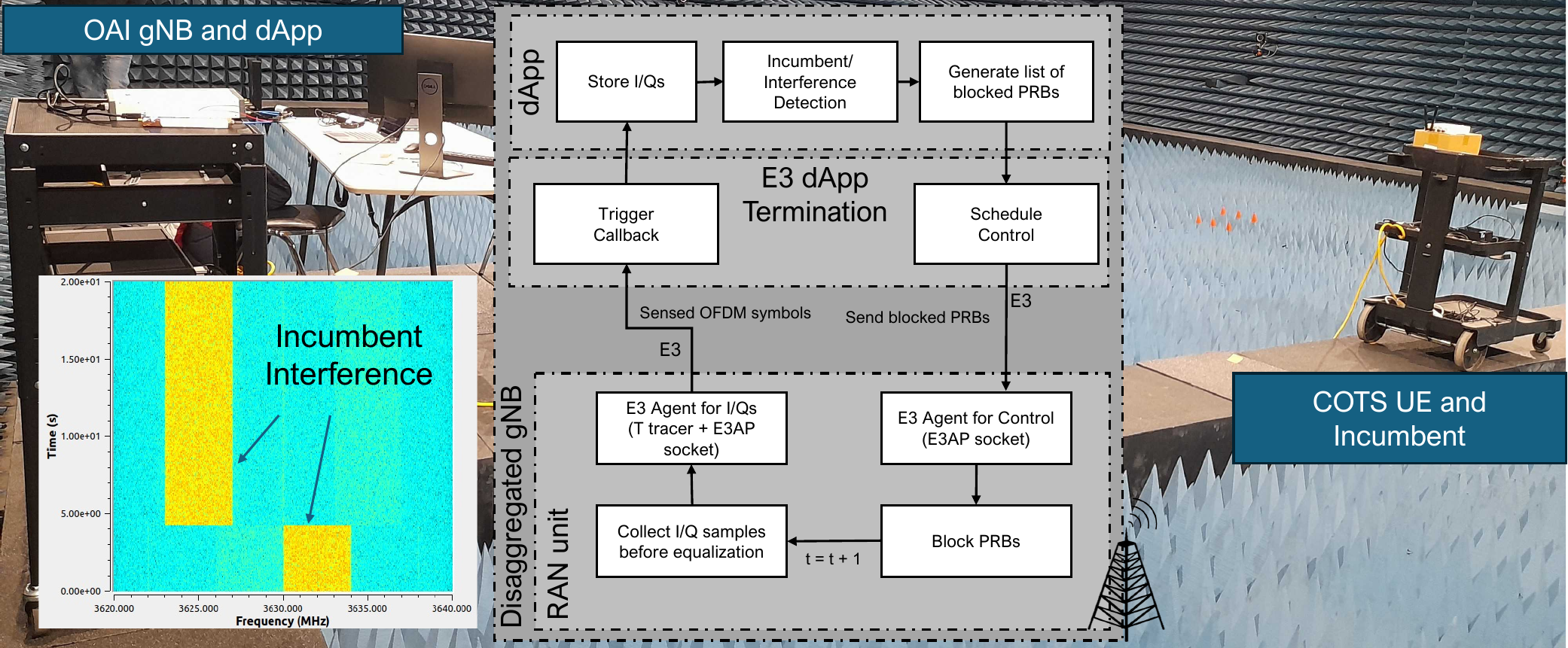}
    \caption{Intelligent real-time control loop for the \gls{prb} blacklisting. We omit in this figure the initial E3\gls{ap} procedures.}
    \label{fig:use-case}
\end{figure*}

Dynamic spectrum sharing among multiple wireless access technologies is expected to play a major role in the next-generation wireless system design.
Currently, a popular spectrum sharing example in the U.S. is the case of the \gls{cbrs} band, where mobile network operators may coexist with high-priority incumbent federal radar and satellite systems users in the 3.55-3.7\,GHz frequency range.
When considering \gls{5g} \gls{nr} systems, the \gls{cbrs} band is a subset of the bands n77 and n78.

In our recent work~\cite{gangula2024listen}, we leveraged our dApp framework to implement a real-time \gls{ran}-driven spectrum sharing system.
In this system, the \gls{gnb} can communicate while simultaneously performing the spectrum sensing task in the \gls{cbrs} band. 
Moreover, the \gls{gnb} can adapt its communication parameters (e.g., frequencies or the list of \glspl{prb} to not be used for scheduling purposes due to presence of high interference) whenever a primary incumbent user is detected.
The spectrum sharing dApp that we developed follows the prototype mentioned in the previous section and illustrated in Fig.~\ref{fig:use-case}. 
The dApp connects with the \gls{gnb} using the E3 and performs the following tasks: 
\begin{enumerate}
    \item Extract I/Q samples from dedicated symbols reserved for spectrum sensing at the \gls{gnb} through a callback registered with the \texttt{E3Interface}, enabling real-time sensing and data extraction;
    \item Leverage an inference algorithm to process the I/Q samples and detect incumbent users by computing the magnitude of the samples and comparing each magnitude with a fixed threshold previously calibrated; 
    \item If an incumbent is detected, create a list of the \glspl{prb} affected by the incumbent and that should not be used by the \gls{gnb} to schedule transmissions;
    \item Deliver such list to the \ran node as a control action through the E3 interface. 
\end{enumerate}

% In our system, we have updated the \gls{oai} \gls{gnb} scheduler so that it avoids allocating these \glspl{prb} to any user.
% This allows for the minimum disruption of cellular services while vacating the spectrum for the incumbent user. 
%
% The processed data can be visualized for debugging or showcasing purposes using an external \gls{gui} implemented in the dApp logic, examples of which are shown in Fig.~\ref{fig:use-case-1}.
% , along with a waterfall plot from a Keysight EXA Spectrum Analyzer N9010B used as external observer.

\begin{figure}[tp]
\centering
\begin{subfigure}[t]{0.5\linewidth}
    \centering
    \includegraphics[width=\linewidth]{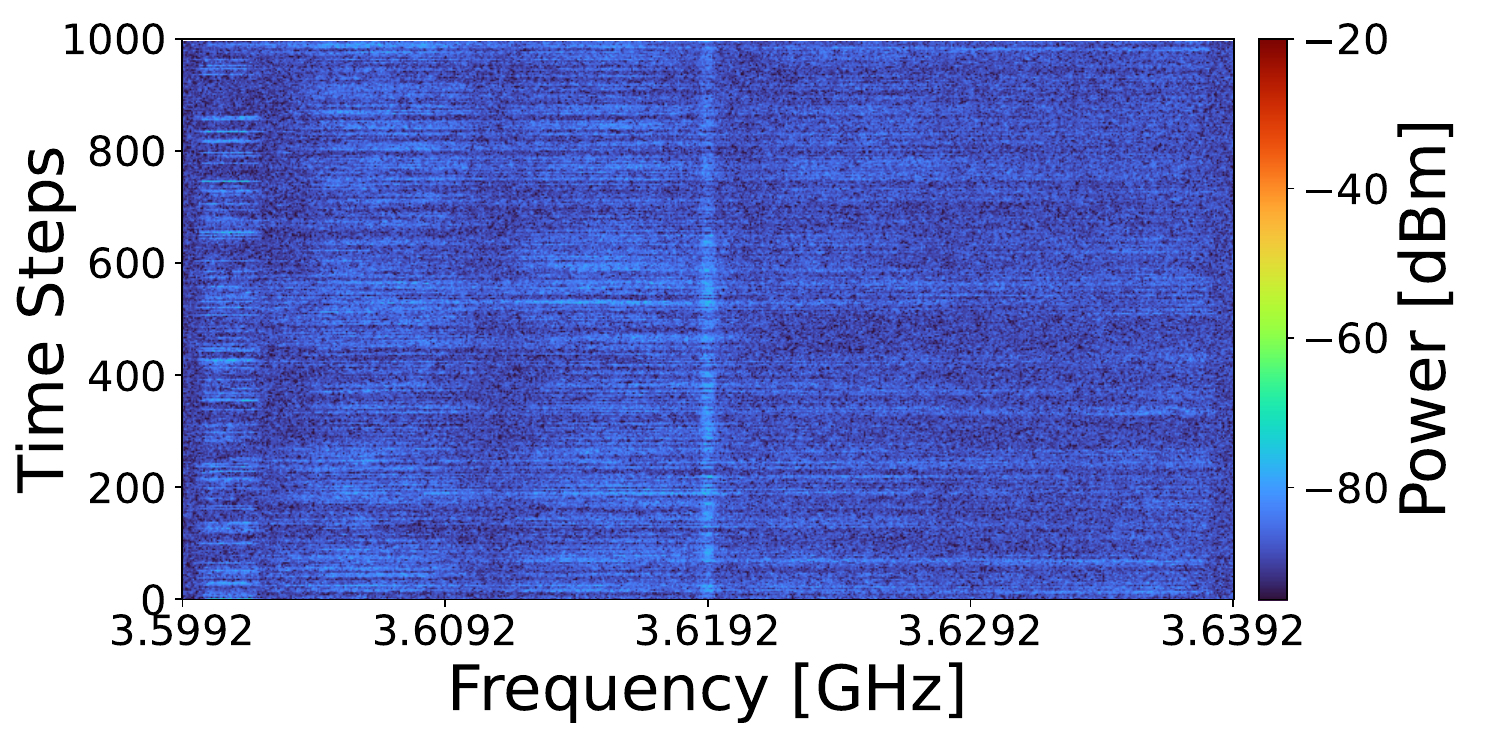}
    \caption{No incumbent. The presence of the dApp here is negligible.}
    \label{fig:5gnormal}
\end{subfigure}\hfil% equal to outside spacing
 \begin{subfigure}[t]{0.5\linewidth}
    \centering
    \includegraphics[width=\linewidth]{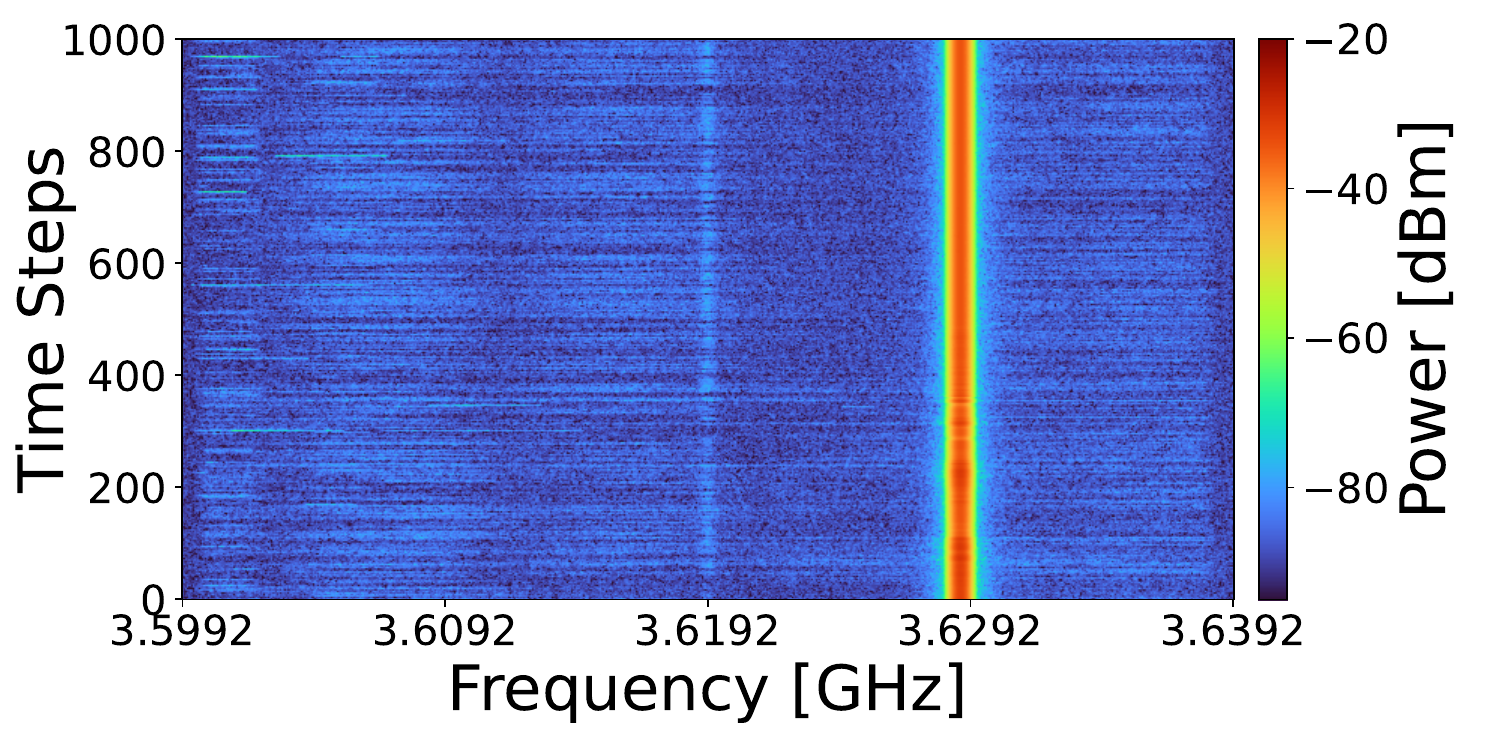}
    \caption{Incumbent clashing with \gls{ue}.}
    \label{fig:5gjammer}
\end{subfigure}\hfil% equal to outside spacing
\begin{subfigure}[t]{0.5\linewidth}
    \centering
    \includegraphics[width=\linewidth]{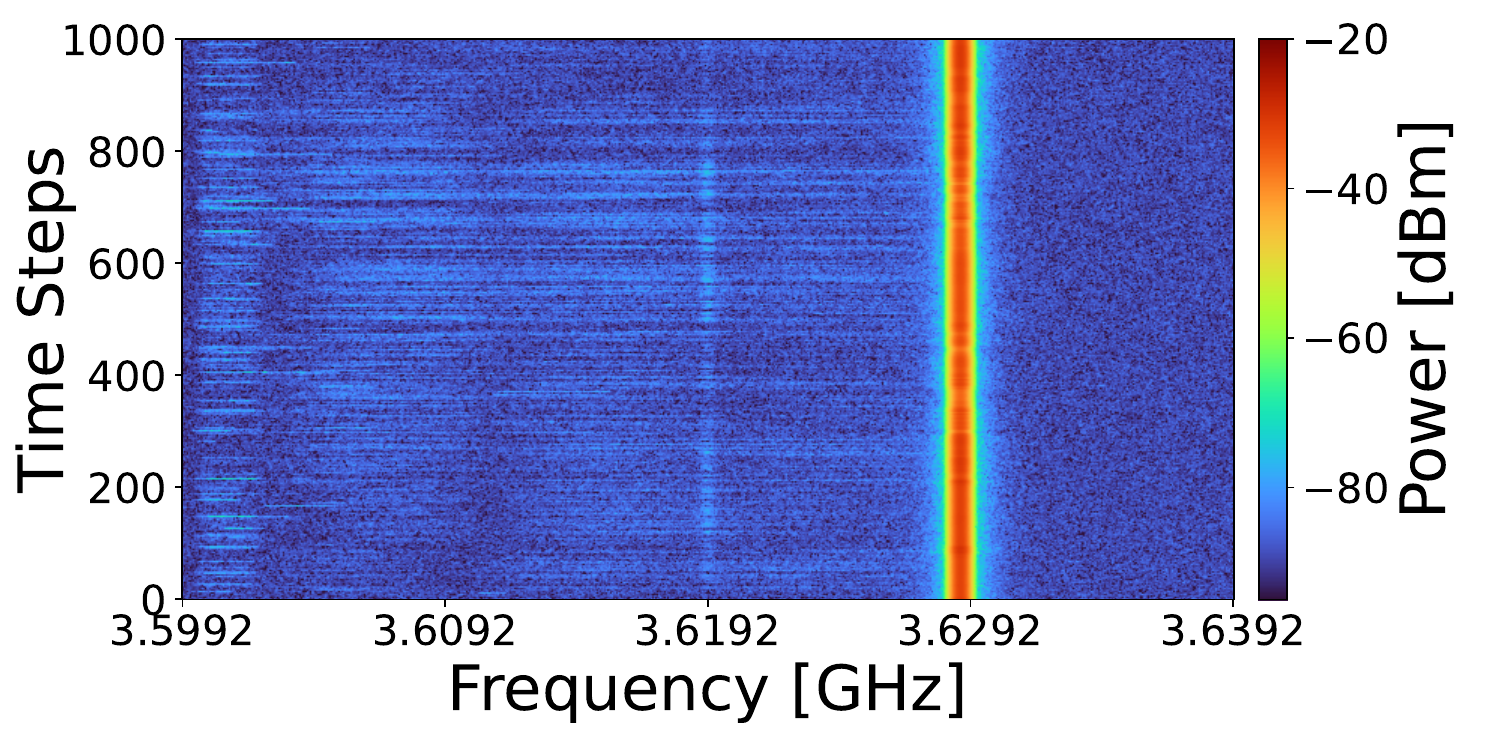}
    \caption[b]{Resulting spectrum when the dApp is blocking the \glspl{prb}.}
    \label{fig:5gdApp}
 \end{subfigure}
  \caption{\textit{Spectrum sharing} dApp cases evaluated in the experiments: Sensed Spectrum of a \gls{5g} \gls{gnb}-\gls{ue} communication with the incumbent invading the spectrum of the \gls{gnb} with and without the dApp.}
 \label{fig:use-case-1}
\end{figure}

% If you want the gui decomment
% \begin{subfigure}[t]{0.5\linewidth}
%     \centering
%     \includegraphics[width=\linewidth]{images/use_case_gui/gui.pdf}
%     \caption[b]{Screenshot of the dApp \gls{gui} showing the sensed spectrum at the \gls{gnb} detecting the presence of an incumbent and the blocking the affected \glspl{prb}}
%     \label{fig:dAppGUI}
%  \end{subfigure}\hfil

We analyzed three different configurations of the spectrum in Fig.~\ref{fig:use-case-1} using a Keysight EXA Spectrum Analyzer N9010B as an external observer.
The first one, shown in Fig.~\ref{fig:5gnormal}, depicts an unbounded \gls{tcp} downlink transmission of a \gls{5g} \gls{gnb} to one \gls{ue}.

In this case, the \gls{mac} scheduler of the \gls{gnb} is allowed to use all possible \glspl{prb}.
Fig.~\ref{fig:5gjammer} introduces a narrowband incumbent in the spectrum of the \gls{gnb},
% without the presence of the Spectrum Sharing dApp,
making the \gls{5g} \gls{rf} signal clash with the incumbent \gls{rf} signal and degrading the \gls{gnb}-\gls{ue} performance.
When the spectrum sharing dApp interacts with the \gls{gnb}, its control logic analyzes the sensed I/Qs and is able to detect the presence of the incumbent and to create a control action to block the clashing \glspl{prb}, making them unavailable to the scheduler. 
% This is shown in Fig.~\ref{fig:dAppGUI}, which depicts the external \gls{gui} of the dApp that reports the data delivered by the \gls{gnb} and the applied control from the dApp.
% Since the reserved symbol for sensing is not allocated to the \gls{ue} traffic by the \gls{gnb}, the I/Qs related to such traffic are not delivered to the dApp.
% In this case, the magnitude of the sensed spectrum presents the incumbent and the radio leakage of the \gls{gnb}, which is anyway not in the space of the \glspl{prb} and thus it is not part of the incumbent detection performed by the dApp.
% Compared to the spectrograms of Figs.~\ref{fig:5gnormal} and~\ref{fig:5gjammer}, the dApp is only receiving and evaluating the sensed symbols from the \gls{gnb}, thus ignoring the actual data messages exchanged with the \gls{ue}.
Finally, the dApp delivers the control action to the \gls{ran}, vacating the affected portion of the spectrum, as shown in Fig.~\ref{fig:5gdApp}.
Under these conditions, \gls{5g} communication can continue to coexist with the incumbent, albeit with certain limitations.

Specifically, we would like to mention that the currently available \gls{oai} scheduler implements a type~1 resource allocation scheme as defined in the \gls{3gpp} \gls{ts} 38.214 Section 6.1.2.2~\cite{3gpp.38.214}, which requires that the \glspl{prb} assigned to a single \gls{ue} form a contiguous, non-interleaved sequence of virtual resource blocks.
As a consequence, when we mute a set of \glspl{prb} already occupied by the incumbent, we notice that the scheduler also blocks all scheduling activities in the \glspl{prb} occupying the rightmost part of the spectrum.
Although this is a limitation that does not pertain our spectrum sensing dApp, it is also fair to point out that this specific implementation based on type~1 scheduler might reduce unnecessarily the list of \glspl{prb} that can be used for data transmission.
However, we also point out that  in the same specifications, precisely in Section~6.1.2.1~\cite{3gpp.38.214}, the standard defines resource allocation of type~0, which allows for the allocation of noncontiguous \glspl{prb} to the same \gls{ue}. 
Unfortunately, this latter type is not implemented in \gls{oai} at the time of this writing, and the evaluation of our dApp under this configuration is left for future studies.

As we will discuss in the next section, the results reported in Table~\ref{tab:use_case} show that, although \gls{oai}'s type~1 scheduler results in an unused portion of the spectrum (right portion of the spectrum in Fig.~\ref{fig:5gdApp}), the usage of a reduced portion of the spectrum does not significantly reduce \gls{ue} downlink performance.
% and to slightly improve them compared to the situation without the dApp, while ensuring spectrum sharing compliance.

\begin{table}[ht]
    \centering
    \begin{tabular}{|c|c|c|c|c|}
    \hline
     \textbf{Testbed} & \textbf{dApp}  &  \textbf{Incumbent} &  \textbf{Throughput (\textit{Mbps})} &  \textbf{Shared Spectrum?} \\ \hline
        Colosseum & N &  N & 71.34 $\pm$ 1.28 & N/A (gNB only)\\ \hline
        Colosseum & Y &  N & 71.53 $\pm$ 0.76 & N/A (gNB only) \\ \hline
        Colosseum & N &  Y & 49.52 $\pm$ 3.37 & \textbf{No coordination}\\ \hline
        Colosseum & Y &  Y & 53.78 $\pm$ 1.55 & \textbf{Shared spectrum}\\ \hline        
        Arena & N &  N & 77.98 $\pm$ 1.31 &  N/A (gNB only)\\ \hline
        Arena & Y &  N & 76.37 $\pm$ 1.87 &  N/A (gNB only) \\ \hline
        Arena & N &  Y & 38.97 $\pm$ 2.95 &  \textbf{No coordination}\\ \hline
        Arena & Y &  Y & 43.86 $\pm$ 1.52 & \textbf{Shared spectrum} \\ \hline
    \end{tabular}
    \caption{Summary of the performance of \gls{oai} and Spectrum Sharing dApp. Average throughput reported with the 95\% confidence interval.}
    \label{tab:use_case}
\end{table}

{\bf Performance Evaluation --} Table~\ref{tab:use_case} summarizes the experiment results of the Spectrum Sharing use case for the two testbeds and the different test configurations considered.
A 95\% confidence interval, using a Z-value of 1.96, was applied to the results and it is reported along with the averages of each measurements. 
Given the radio configuration of the experiments presented in Sec.\ref{sec:loop_benchmark}, the incumbent may interfere with the \gls{gnb} spectrum if transmitting between 3.5992\,GHz and 3.6392\,GHz. 
We present the throughput results across various configurations: with and without the incumbent, and with and without the spectrum sharing dApp. These results highlight the scenarios where spectrum is shared and where it is not.
For these experiments, the incumbent signal has been created using the \texttt{uhd\_siggen} software set up to generate an uniform noise transmitted at 3.63\,GHz with a sampling rate of 1\,MHz, an amplitude of 0.5\,and gain of 90\,dB on Arena and 60\,dB on Colosseum.
A dBm threshold used to determine whether to block the \glspl{prb} was calibrated specifically for each experimental environment.
It is worth mentioning that each testbed is characterized by different radios with diverse sensitivity, and different RF conditions. For this reason, we set the noise floor for the two testbeds considered in our analysis to two different value. Specifically, we set the noise floor to 20\,dBm and 53\,dBm for Arena and Colosseum, respectively.
% In Arena, a threshold of 20\,dBm was used, while a 53\,dBm one was used in Colosseum, as its noise floor is higher that than in the Arena testbed.%\hl{cite Davide's paper on noise floor, is there one?}.

The choice of these values is conservative. Indeed, a too high of a threshold value might fail in detecting incumbent activities in low \gls{sinr} regimes, and might result in \gls{gnb} transmissions that overlap with the incumbent and cause severe interference. Similarly, a too low of a threshold might misinterpret noise as incumbent signals, and block all \glspl{prb}. In our case, we have evaluated the thresholds heuristically via experiments and data analytics. Note that, in general, a conservative threshold is to be preferred as incumbents have always priority and \glspl{gnb} must vacated promptly the spectrum.
% Consequently, the dApp blocks a larger bandwidth, resulting in a higher number of blocked \glspl{prb}, which amplifies the performance degradation of the \gls{5g} network.

Results in Table \ref{tab:use_case} show that the execution of the dApp (cases where dApp = Y and Incumbent = N) does not introduce any significant impact on throughput if compared to the case where the dApp is not enabled (cases where dApp = N and Incumbent = N). %\hl{provide numbers, say that despite small variations, magnitude is pretty much the same}
Specifically, in the Colosseum testbed, the average throughput measured without the dApp is 71.34\,Mbps, while we measure an average throughput of 71.53\,Mbps when the dApp is enabled. 
Similarly, in the Arena testbed, the throughput decreases slightly from 77.98\,Mbps without the dApp to 76.37\,Mbps with the dApp. 
% Despite the average values seem to suggest a performance degradation, we notice that confidence intervals reported in Table~\ref{tab:use_case} suggest that the throughput remains within the same order of magnitude, demonstrating that the dApp introduces minimal overhead in these scenarios that allows for performance parity.
%
\new{Although the average values may suggest performance degradation, the confidence intervals in Table~\ref{tab:use_case} indicate that throughput takes values within the same confidence range, demonstrating that the dApp introduces minimal overhead and maintains performance parity.}

In the case where the Incumbent is present (i.e., when Incumbent = Y), the \gls{rf} signals of \gls{gnb} and incumbent interfere with each other.
The lack of spectrum sharing (cases where dApp = N and Incumbent = Y) causes the gNB to use \glspl{prb} affected by interference, which significantly degrades the \gls{ue} throughput by nearly 50\% across both testbeds.
% For this rteason Finally, we evaluate the case in which both dApp and Incumbent are active at the same time, i.e., .
We instead notice that the presence of the spectrum sharing dApp (cases where dApp = Y and Incumbent = Y) is beneficial, as the blocking of \glspl{prb} affected by interference allows the sharing of a portion of the spectrum with the incumbent while preventing disruptive interference. This brings a slight improvement of the \gls{ue} throughput because preventing the use of interfered \glspl{prb} results in less interference, lower errors and reduced need for retransmissions.

\subsection{Use Case: Sensing and Positioning}
\label{sec:use-case-2}

\begin{figure}[ht]
\centering
    \includegraphics[scale=0.4]{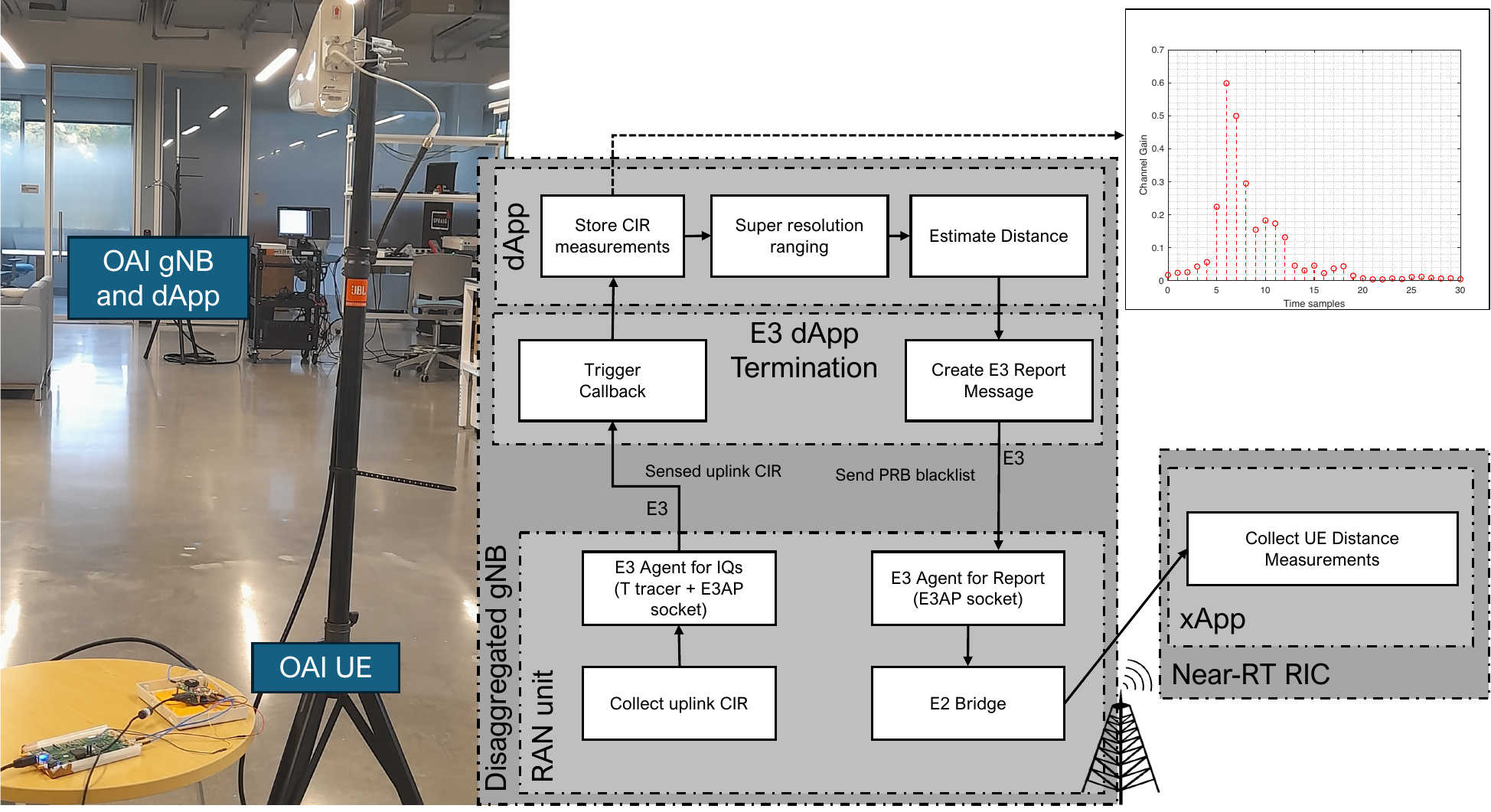}
    \caption{Setup for the Sensing and Positioning dApp and UL CIR extracted by the dApp.}
    \label{fig:use-case-2-scenario}
\end{figure}

Cellular operations at higher frequencies enable the use of wider bandwidths (e.g., \gls{fr2}, \gls{fr3}, and THz) and massive antenna arrays. Thanks to these technologies, beyond 5G systems are expected to offer not only higher data rates but also high-resolution positioning and environment sensing capabilities~\cite{bourdoux20206g}.
%In the current \gls{3gpp} positioning framework, the \gls{lmf} entity within the 5G core network oversees location services. It coordinates the collection of positioning related measurements and is responsible for calculating the location of the \gls{ue}.
In current \gls{ul}-based positioning methods,
% within the \gls{3gpp} framework,
the \gls{gnb} derives specific positioning-related measurements (defined in \gls{3gpp} standards) from the \gls{ul} channel estimates and forwards them to the \gls{lmf} entity within the 5G core network.
These measurements can be encoded into a few bits and are generally derived from low-complexity signal processing algorithms to comply with the real-time nature of the \gls{gnb}. 
%Moreover, these algorithms work on few channel measurements.

On the other hand, super-resolution and \gls{ml}-based algorithms offer precise sensing and positioning capabilities that outperform conventional algorithms~\cite{Nessa2020mlpos,xinrong2004}.
% However, they are data intensive and complex, complicating their implementation within the \gls{gnb} protocol stack.
However, they usually require data that is not accessible from outside the \ran, such as the \gls{cir}.
Therefore, dApps are a natural tool to address this problem. 

In our recent work~\cite{gangula2024round}, we have used the dApp framework in an \gls{ul} ranging/positioning system, where the distance between the \gls{ue} and \gls{gnb} is computed based on the \gls{ue}'s \gls{ul} \gls{cir} available at the \gls{gnb} and exposed to the dApp.
To demonstrate this method, we have developed a testbed consisting of an \gls{oai} \gls{gnb} (enhanced with the E3 agent), a ranging dApp, and \gls{oai} \gls{ue}.
The dApp connects with the \gls{oai} \gls{gnb} using the E3 agent and performs the following tasks: 
(i) extracts multiple wideband \gls{ul} \gls{cir} measurements of a \gls{ue} from the \gls{gnb}; and (ii) runs a super-resolution algorithm~\cite{gangula2024round} on the collected \gls{cir} measurements to find the distance between the \gls{ue} and the \gls{gnb}.

\begin{figure}[ht]
\centering
    \includegraphics[scale=0.8]{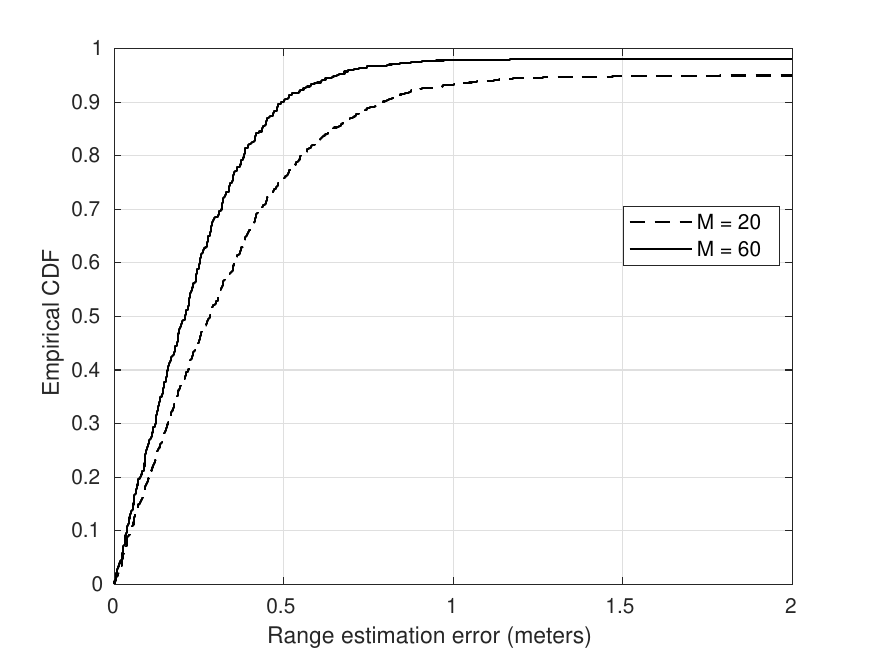}
    \caption{\gls{cdf} of the range estimation error.}
    \label{fig:range_error}
\end{figure}

\new{The above described ranging dApp framework is tested and validated with real-world experiments. 
The experimental setup consists of a single antenna \gls{gnb} and a \gls{ue} communicate over a line-of-sight channel. The \gls{gnb} and the \gls{ue} rely on on the \gls{oai} protocol stack and \gls{usrp} B210 software-defined boards. Furthermore, the SC2430 \gls{nr} signal conditioning module is used as an external \gls{rf} front-end at the \gls{gnb} \cite{sc_mod}. 
The system operates at band n78 with a carrier frequency of $3.69$\,GHz and a bandwidth of 40\,MHz.}

\new{
{\bf Performance Evaluation --} 
The performance of the proposed ranging scheme is evaluated 
in terms of empirical \gls{cdf} of the distance estimation error at an \gls{ul} \gls{snr} of $-20$ dB.
The CDF is obtained from 48,000 \gls{cir} measurements by fixing the position of the \gls{gnb} and moving the \gls{ue} in a straight line between 3 to 10 meters with a 1-meter increment. At every point, a total of 6,000 measurements are collected.
The \textit{ranging dApp} is used to extract these measurements as shown in Fig.~\ref{fig:use-case-2-scenario}. The \gls{cdf} of the range estimation error with a super resolution \gls{music} algorithm 
with $M{=}20$ to $60$ number of \gls{cir} measurements is shown in Figure~\ref{fig:range_error}.
For details on ranging algorithms and prototype design, we refer the readers to~\cite{gangula2024round}.
While the \textit{ranging dApp} here is used to collect measurements during the experiment, which are then processed offline to obtain the range error CDF, it's important to highlight that it can also be used for real-time \gls{cir} collection and distance evaluation. The measurement collection latency performance is similar to that described in Section \ref{sec:loop_benchmark}.
}
%The system operates at band n78 with a carrier frequency of $3.69$\,GHz and a bandwidth of 40\,MHz. The \gls{ul} \gls{cir} extracted by the \textit{ranging dApp} is shown in Fig.~\ref{fig:use-case-2-scenario}.

%The channel gain and channel impulse response both depend on \gls{ue}’s position because path loss, shadowing, and multi-path are all phenomena that vary with distance and location of the \gls{ue}. Closer \glspl{ue} benefit from lower path loss and thus higher gain, while changes in nearby obstacles and reflections impact the channel impulse response by altering the number and timing of multi-path components. 
%As a \gls{ue} moves, both large-scale fading (due to environmental obstructions) and small-scale fading (due to the interference of different signal paths) continually vary the channel characteristics, causing the gain and impulse response to change accordingly. For details on ranging algorithms and prototype design, we refer the readers to~\cite{gangula2024round}.

%An over-the-air indoor experimental setup with the \gls{oai} \gls{gnb} and \gls{ue} along with the \gls{ul} \gls{cir} extracted by the \textit{ranging dApp} is shown in Fig.~\ref{fig:use-case-2-scenario}. The system operates at a carrier frequency of $3.69$\,GHz with a bandwidth of 40\,MHz. 

\section{Conclusions}
\label{sec:conclusions}

In this paper, we introduced an extension of the \oran architecture focused on enabling (i) real-time control in the \gls{ran}, and (ii) the interaction of the \oran stack with the user-plane of the network. This is achieved through dApps, lightweight microservices co-located with \glspl{du} and \glspl{cu}, and holistically managed through components such as the \glspl{ric} and \gls{smo}. We first described the role that dApps have in enabling new use cases for \gls{ran} optimization, including inference based on I/Q samples, reference signals available at the physical layer, \new{or as complex \gls{isac} systems}. Based on the data and control requirements defined by these use cases, we discussed the architecture for the integration of dApps and \gls{ran} nodes. We also provided insights on the design and implementation of a set of \glspl{api} that \gls{ran} nodes need to expose to dApps, coordinated by an \gls{api} broker called E3 agent. We then discussed the \gls{lcm} for dApps, including steps for development, onboarding, and deployment of dApps, as well as procedures for interaction with the \glspl{ric} and \gls{smo}.

Then, we proposed a reference implementation for dApps based on \gls{oai}, which is publicly available to enable research and development of applications and use cases relying on real-time control loops in Open \ran.
We described the dApp framework, based on Python, and the integration with \gls{oai}. We then presented two use cases based on our reference implementation. For the spectrum sharing one, a dApp performs spectrum sensing at the \gls{du},
% as a spectrum sensor,
understands which portions of the spectrum are occupied by incumbents, and coordinates with the \gls{du} scheduler to avoid scheduling over it. 
In the analysis of this use case, results show that while the presence of a continuous incumbent affects the performance of the \gls{5g} network, the use of a dApp enables the sharing of the spectrum bringing a slight improvement of the \gls{5g} network performance.
% (mostly blocked by the standard scheduler and not by the operations of the dApp itself).
% The implementation of the resource allocation type~0 in \gls{oai} will most likely improve the results of the dApp reported in this work without changing the methodology presented in the paper.
% Moreover, we acknowledge that a dynamic calibration of the noise floor instead of the fixed threshold used in our experiment would further improve the throughput, but in the context of this work, we opted for a conservative solution that still improves the throughput while performing \gls{rf} spectrum sharing.

For the positioning use case, the dApp processes uplink reference signals to extract the uplink channel response and perform ranging for the \gls{ue}.
We profiled the performance of our dApp framework, demonstrating that real-time control loops under 1\,ms are achievable in \oran. In our implementation, the average control loop duration is slightly less than 400\,$\mu$s, on average.

% Finally, we compared dApps with other approaches, including real-time \glspl{ric}, and concluded the paper. 
As part of our future work, we are extending the dApp framework to other open stacks, including NVIDIA ARC-OTA~\cite{villa2024x5g}, and integrate the dApps in a continuous integration, deployment, and testing framework to continuously test it on and up-to-date \gls{gnb} protocol stack.

\appendix % Start of the appendix section

\section{dApp Deployment Process}
\label{appendix:dapp-deployment}
The following table provides a detailed breakdown of the \gls{lcm} process for the dApp, outlining the steps, roles, and conditions involved in its deployment and operation.
\begin{longtable}{|>{\raggedright\arraybackslash}p{0.25\linewidth}|>{\raggedright\arraybackslash}p{0.65\linewidth}|} 
    \caption{dApp Deployment} \label{tab:lcm}\\
    \hline
    \multicolumn{2}{|c|}{\textbf{Use Case Stages and Evolution/Specification}} \\
    \hline
    \textbf{Goal} & To instantiate and configure a dApp on theCU/DU, facilitating its deployment, management, and operationalization directly via the SMO and the CU/DU. \\
    \hline
    \textbf{Actors and Roles} & 
    \begin{itemize}
        \item \textbf{SMO (Service Management and Orchestration)}: Orchestrates the deployment, configuration, and lifecycle management of the dApp.
        \item \textbf{DU (Distributed Unit)}: Possible host of the dApp and provides necessary network and compute resources for its operation.
        \item \textbf{CU (Central Unit)}: Possible host the dApp and provides necessary network and compute resources for its operation.
        \item \textbf{Network Function Orchestration (NFO)}: Interfaces with the DU for deploying and configuring the dApp.
        \item \textbf{Deployment Management Services (DMS)}: Manages the deployment resources on the DU.
    \end{itemize} \\
    \hline
    \textbf{Assumptions} & 
    \begin{itemize}
        \item The SMO and CU/DU are available and operational.
        \item The dApp package has been validated, verified, and cataloged in the SMO's runtime library.
        \item The DU is pre-configured and ready to host dApp instances.
    \end{itemize} \\
    \hline
    \textbf{Preconditions} & 
    \begin{itemize}
        \item The dApp is onboarded and certified within the SMO.
        \item The CU/DU has the necessary resources and access to the container images required for the dApp deployment.
    \end{itemize} \\
    \hline
    \textbf{Begins When} & A Network Function Install Project Manager initiates a request to the SMO for deploying a new dApp instance on the CU/DU. \\
    \hline
    \textbf{Step 1 (M)} & The SMO receives a service request to deploy the dApp instance on the CU/DU. \\
    \hline
    \textbf{Step 2 (M)} & The SMO decomposes the service request and identifies all dApps to be deployed and their deployment order. \\
    \hline
    \textbf{Step 3 (M)} & The SMO determines which DU, or CU, and deployment parameters to use. This is based on policies or explicit input from the Network Function Install Project Manager. \\
    \hline
    \textbf{Step 4 (M)} & The SMO retrieves the CloudNativeDescriptor for the dApp from the runtime library. \\
    \hline
    \textbf{Step 5 (M)} & The SMO directs the O-Cloud DMS using O2 to create the dApp deployment. \\
    \hline
    \textbf{Step 6 (M)} & DMS allocates the necessary compute, storage, and network resources on the CU/DU as per the dApp deployment request. \\
    \hline
    \textbf{Step 7 (M)} & The SMO sets up the initial configuration for the dApp, such as environment variables, network policies, and access parameters. \\
    \hline
    \textbf{Step 8 (M)} & DMS deploys the dApp container(s) on the CU/DU, setting up the necessary resources and connectivity. \\
    \hline
    \textbf{Step 9 (M)} & DMS notifies the SMO that the dApp deployment has been successfully instantiated and provides a Deployment ID. \\
    \hline
    \textbf{Step 10 (M)} & The SMO updates its dApp inventory with the new Deployment ID and deployment status. \\
    \hline
    \textbf{Step 11 (M)} & The deployed dApp instance reads its initial configuration from the provided parameters and begins its operation. \\
    \hline
    \textbf{Step 12 (O)} & The SMO continuously monitors the dApp’s health, performance, and connectivity. It can also perform scaling, updates, or redeployment as required. \\
    \hline
    \textbf{Step 13 (M)} & The SMO informs the Network Function Install Project Manager of the overall success or failure of the request. \\
    \hline
    \textbf{Ends When} & The dApp instance is successfully deployed, configured, and operational on the CU/DU, with the SMO actively managing its lifecycle. \\
    \hline
    \textbf{Post Conditions} & 
    \begin{itemize}
        \item The dApp is actively running and functional on the CU/DU.
        \item The SMO has an updated inventory reflecting the deployed dApp, and lifecycle management is in place.
    \end{itemize} \\
    \hline
    \textbf{Exceptions} & If the deployment fails, the DMS notifies the SMO, and the SMO informs the Network Function Install Project Manager to take corrective actions. \\
    \hline
\end{longtable}

\section*{Acknowledgement}

This article is based upon work partially supported by OUSD(R\&E) through Army Research Laboratory Cooperative Agreement Number W911NF-24-2-0065. The views and conclusions contained in this document are those of the authors and should not be interpreted as representing the official policies, either expressed or implied, of the Army Research Laboratory or the U.S. Government. The U.S. Government is authorized to reproduce and distribute reprints for Government purposes notwithstanding any copyright notation herein. The work was also partially supported by SERICS (PE00000014) 5GSec project, CUP B53C22003990006, under the MUR National Recovery and Resilience Plan funded by the European Union - NextGenerationEU, and by the U.S.\ National Science Foundation under grants CNS-1925601, CNS-2117814, and CNS-2112471.

\footnotesize  % for natbib
\bibliographystyle{IEEEtran}
\bibliography{biblio}

\begin{thebibliography}{10}
\providecommand{\url}[1]{#1}
\csname url@samestyle\endcsname
\providecommand{\newblock}{\relax}
\providecommand{\bibinfo}[2]{#2}
\providecommand{\BIBentrySTDinterwordspacing}{\spaceskip=0pt\relax}
\providecommand{\BIBentryALTinterwordstretchfactor}{4}
\providecommand{\BIBentryALTinterwordspacing}{\spaceskip=\fontdimen2\font plus
\BIBentryALTinterwordstretchfactor\fontdimen3\font minus
  \fontdimen4\font\relax}
\providecommand{\BIBforeignlanguage}[2]{{%
\expandafter\ifx\csname l@#1\endcsname\relax
\typeout{** WARNING: IEEEtran.bst: No hyphenation pattern has been}%
\typeout{** loaded for the language `#1'. Using the pattern for}%
\typeout{** the default language instead.}%
\else
\language=\csname l@#1\endcsname
\fi
#2}}
\providecommand{\BIBdecl}{\relax}
\BIBdecl

\bibitem{johnson2022nexran}
D.~Johnson, D.~Maas, and J.~Van Der~Merwe, ``{NexRAN: Closed-loop RAN slicing
  in POWDER-A top-to-bottom open-source open-RAN use case},'' in
  \emph{Proceedings of the 15th ACM Workshop on Wireless Network Testbeds,
  Experimental evaluation \& CHaracterization}, 2022, pp. 17--23.

\bibitem{polese2022colo}
M.~Polese, L.~Bonati, S.~D'Oro, S.~Basagni, and T.~Melodia, ``Col{O}-{RAN}:
  {D}eveloping machine learning-based x{A}pps for open {RAN} closed-loop
  control on programmable experimental platforms,'' \emph{IEEE Transactions on
  Mobile Computing}, vol.~22, no.~10, pp. 5787--5800, 2022.

\bibitem{lacava2024programmable}
A.~Lacava, M.~Polese, R.~Sivaraj, R.~Soundrarajan, B.~S. Bhati, T.~Singh,
  T.~Zugno, F.~Cuomo, and T.~Melodia, ``Programmable and {C}ustomized
  {I}ntelligence for {T}raffic {S}teering in {5G} {N}etworks {U}sing {O}pen
  {RAN} {A}rchitectures,'' \emph{IEEE Transactions on Mobile Computing},
  vol.~23, no.~4, pp. 2882--2897, 2024.

\bibitem{dryjanski2021toward}
M.~Dryja{\'n}ski, {\L}.~Ku{\l}acz, and A.~Kliks, ``Toward modular and flexible
  open ran implementations in 6g networks: Traffic steering use case and o-ran
  xapps,'' \emph{Sensors}, vol.~21, no.~24, p. 8173, 2021.

\bibitem{10620740}
K.~Suzuki, J.~Nakazato, Y.~Sasaki, K.~Maruta, M.~Tsukada, and H.~Esaki,
  ``{T}oward {B5G/6G} {C}onnected {A}utonomous {V}ehicles: {O-RAN-D}riven
  {M}illimeter-{W}ave {B}eam {M}anagement and {H}andover {M}anagement,'' in
  \emph{IEEE Conference on Computer Communications Workshops (INFOCOM WKSHPS)},
  2024, pp. 1--6.

\bibitem{kundu2024towards}
L.~Kundu, X.~Lin, and R.~Gadiyar, ``{T}owards {E}nergy {E}fficient {RAN}:
  {F}rom {I}ndustry {S}tandards to {T}rending {P}ractice,'' \emph{arXiv
  preprint arXiv:2402.11993}, 2024.

\bibitem{bacsaran2023deep}
O.~T. Ba{\c{s}}aran, M.~Ba{\c{s}}aran, D.~Turan, H.~G. Bayrak, and Y.~S.
  Sandal, ``{D}eep {A}utoencoder {D}esign for {RF} {A}nomaly {D}etection in
  {5G} {O-RAN} {N}ear-{RT} {RIC} via x{A}pps,'' in \emph{2023 IEEE
  International Conference on Communications Workshops (ICC Workshops)}.\hskip
  1em plus 0.5em minus 0.4em\relax IEEE, 2023, pp. 549--555.

\bibitem{10356316}
A.~Tripathi, J.~S.~R. Mallu, M.~H. Rahman, A.~Sultana, A.~Sathish, A.~Huff,
  M.~Roy~Chowdhury, and A.~P. Da~Silva, ``{E}nd-to-{E}nd {O-RAN}
  {C}ontrol-{L}oop {F}or {R}adio {R}esource {A}llocation in {SDR}-{B}ased {5G}
  {N}etwork,'' in \emph{IEEE Military Communications Conference (MILCOM)},
  2023, pp. 253--254.

\bibitem{al2020exposing}
A.~Al-Shawabka, F.~Restuccia, S.~D’Oro, T.~Jian, B.~C. Rendon, N.~Soltani,
  J.~Dy, S.~Ioannidis, K.~Chowdhury, and T.~Melodia, ``{E}xposing the
  {F}ingerprint: {D}issecting the {I}mpact of the {W}ireless {C}hannel on
  {R}adio {F}ingerprinting,'' in \emph{IEEE INFOCOM 2020-IEEE Conference on
  Computer Communications}.\hskip 1em plus 0.5em minus 0.4em\relax IEEE, 2020,
  pp. 646--655.

\bibitem{9869746}
S.-D. Wang, H.-M. Wang, C.~Feng, and V.~C.~M. Leung, ``{S}equential {A}nomaly
  {D}etection {A}gainst {D}emodulation {R}eference {S}ignal {S}poofing in {5G}
  {NR},'' \emph{IEEE Transactions on Vehicular Technology}, vol.~72, no.~1, pp.
  1291--1295, 2023.

\bibitem{uvaydov2021deepsense}
D.~Uvaydov, S.~D’Oro, F.~Restuccia, and T.~Melodia, ``{D}eepsense: {F}ast
  {W}ideband {S}pectrum {S}ensing {T}hrough {R}eal-{T}ime {I}n-the-{L}oop
  {D}eep {L}earning,'' in \emph{IEEE INFOCOM 2021-IEEE Conference on Computer
  Communications}.\hskip 1em plus 0.5em minus 0.4em\relax IEEE, 2021, pp.
  1--10.

\bibitem{chen2008toward}
R.~Chen, J.-M. Park, Y.~T. Hou, and J.~H. Reed, ``{T}oward {S}ecure
  {D}istributed {S}pectrum {S}ensing in {C}ognitive {R}adio {N}etworks,''
  \emph{IEEE Communications Magazine}, vol.~46, no.~4, pp. 50--55, 2008.

\bibitem{liu2022integrated}
F.~Liu, Y.~Cui, C.~Masouros, J.~Xu, T.~X. Han, Y.~C. Eldar, and S.~Buzzi,
  ``Integrated sensing and communications: Toward dual-functional wireless
  networks for 6g and beyond,'' \emph{IEEE journal on selected areas in
  communications}, vol.~40, no.~6, pp. 1728--1767, 2022.

\bibitem{qadir2023towards}
Z.~Qadir, K.~N. Le, N.~Saeed, and H.~S. Munawar, ``{T}owards {6G} {I}nternet of
  {T}hings: {R}ecent advances, use cases, and open challenges,'' \emph{ICT
  express}, vol.~9, no.~3, pp. 296--312, 2023.

\bibitem{d2022dapps}
S.~D'Oro, M.~Polese, L.~Bonati, H.~Cheng, and T.~Melodia, ``{dApps: Distributed
  applications for real-time inference and control in O-RAN},'' \emph{IEEE
  Communications Magazine}, vol.~60, no.~11, pp. 52--58, 2022.

\bibitem{ko2024edgeric}
W.-H. Ko, U.~Ghosh, U.~Dinesha, R.~Wu, S.~Shakkottai, and D.~Bharadia,
  ``{EdgeRIC: Empowering Real-time Intelligent Optimization and Control in
  NextG Cellular Networks},'' in \emph{21st USENIX Symposium on Networked
  Systems Design and Implementation (NSDI 24)}, 2024, pp. 1315--1330.

\bibitem{upadhyaya2023open}
P.~S. Upadhyaya, N.~Tripathi, J.~Gaeddert, and J.~H. Reed, ``{Open AI Cellular
  (OAIC): An Open Source 5G O-RAN Testbed for Design and Testing of AI-Based
  RAN Management Algorithms},'' \emph{IEEE Network}, vol.~37, no.~5, pp. 7--15,
  2023.

\bibitem{liu2023tinyric}
C.~Liu, G.~Aravinthan, A.~Kak, and N.~Choi, ``{TinyRIC: Supercharging O-RAN
  Base Stations with Real-time Control},'' in \emph{Proceedings of the 29th
  Annual International Conference on Mobile Computing and Networking}, ser. ACM
  MobiCom '23.\hskip 1em plus 0.5em minus 0.4em\relax New York, NY, USA:
  Association for Computing Machinery, 2023.

\bibitem{dappsOranReport}
\BIBentryALTinterwordspacing
{Northeastern University}, NVIDIA, Mavenir, MITRE, and Qualcomm, ``{dApps for
  Real-Time RAN Control: Use Cases and Requirement},'' {O-RAN next Generation
  Research Group (nGRG)}, Research Report, 10 2024, report ID: RR-2024-10.
  [Online]. Available:
  \url{https://mediastorage.o-ran.org/ngrg-rr/nGRG-RR-2024-10-dApp%20use%20cases%20and%20requirements.pdf}
\BIBentrySTDinterwordspacing

\bibitem{polese2024colosseum}
M.~Polese, L.~Bonati, S.~D’Oro, P.~Johari, D.~Villa, S.~Velumani, R.~Gangula,
  M.~Tsampazi, C.~P. Robinson, G.~Gemmi \emph{et~al.}, ``{Colosseum: The Open
  RAN Digital Twin},'' \emph{IEEE Open Journal of the Communications Society},
  2024.

\bibitem{BERTIZZOLO2020107436}
\BIBentryALTinterwordspacing
L.~Bertizzolo, L.~Bonati, E.~Demirors, A.~Al-shawabka, S.~D’Oro,
  F.~Restuccia, and T.~Melodia, ``{Arena: A 64-antenna SDR-based Ceiling Grid
  Testing Platform for Sub-6GHz 5G-and-Beyond radio Spectrum Research},''
  \emph{Computer Networks}, vol. 181, p. 107436, 2020. [Online]. Available:
  \url{https://www.sciencedirect.com/science/article/pii/S1389128620311257}
\BIBentrySTDinterwordspacing

\bibitem{polese2023understanding}
M.~Polese, L.~Bonati, S.~D’oro, S.~Basagni, and T.~Melodia, ``{Understanding
  O-RAN: Architecture, Interfaces, Algorithms, Security, and Research
  Challenges},'' \emph{IEEE Communications Surveys \& Tutorials}, 2023.

\bibitem{abdalla2022toward}
A.~S. Abdalla, P.~S. Upadhyaya, V.~K. Shah, and V.~Marojevic, ``{Toward Next
  Generation Open Radio Access Networks: What O-RAN Can and Cannot Do!}''
  \emph{IEEE Network}, vol.~36, no.~6, pp. 206--213, 2022.

\bibitem{foukas2023taking}
X.~Foukas, B.~Radunovic, M.~Balkwill, and Z.~Lai, ``{Taking 5G RAN Analytics
  and Control to a New Level},'' in \emph{Proceedings of the 29th Annual
  International Conference on Mobile Computing and Networking}, ser. ACM
  MobiCom '23.\hskip 1em plus 0.5em minus 0.4em\relax New York, NY, USA:
  Association for Computing Machinery, 2023.

\bibitem{maxenti2024scaloran}
S.~Maxenti, S.~D'Oro, L.~Bonati, M.~Polese, A.~Capone, and T.~Melodia,
  ``{ScalO-RAN: Energy-aware Network Intelligence Scaling in Open RAN},'' in
  \emph{Proc. of IEEE Intl. Conf. on Computer Communications (INFOCOM)}, May
  2024.

\bibitem{charm}
L.~Baldesi, F.~Restuccia, and T.~Melodia, ``{ChARM: NextG Spectrum Sharing
  Through Data-Driven Real-Time O-RAN Dynamic Control},'' in \emph{IEEE INFOCOM
  2022 - IEEE Conference on Computer Communications}, 2022, pp. 240--249.

\bibitem{villa2023twinning}
D.~Villa, D.~Uvaydov, L.~Bonati, P.~Johari, J.~M. Jornet, and T.~Melodia,
  ``{Twinning Commercial Radio Waveforms in the Colosseum Wireless Network
  Emulator},'' in \emph{Proceedings of the 17th ACM Workshop on Wireless
  Network Testbeds, Experimental evaluation \& Characterization}, 2023, pp.
  33--40.

\bibitem{polese2021deepbeam}
M.~Polese, F.~Restuccia, and T.~Melodia, ``{DeepBeam: Deep Waveform Learning
  for Coordination-Free Beam Management in mmWave Networks},'' in
  \emph{Proceedings of the Twenty-second International Symposium on Theory,
  Algorithmic Foundations, and Protocol Design for Mobile Networks and Mobile
  Computing}, 2021, pp. 61--70.

\bibitem{tractor}
J.~Groen, M.~Belgiovine, U.~Demir, B.~Kim, and K.~Chowdhury, ``{TRACTOR:
  Traffic Analysis and Classification Tool for Open RAN},'' in \emph{ICC 2024 -
  IEEE International Conference on Communications}, 2024, pp. 4894--4899.

\bibitem{oran-wg3-e2-sm-llc}
{O-RAN Working Group 3}, ``{O-RAN} {E2} {S}ervice {M}odel ({E2SM}), {L}ower
  {L}ayers control 1.0,'' ORAN-WG3.TS.E2SM-LLC-R004-v01.00 Technical
  Specification, 2 2025.

\bibitem{m3mimoJsacPaper}
Y.~Chen, Y.~T. Hou, W.~Lou, J.~H. Reed, and S.~Kompella, ``{M3: A
  Sub-Millisecond Scheduler for Multi-Cell MIMO Networks under C-RAN
  Architecture},'' in \emph{IEEE INFOCOM 2022 - IEEE Conference on Computer
  Communications}, 2022, pp. 130--139.

\bibitem{qualcommSharing}
\BIBentryALTinterwordspacing
Qualcomm and {Dell Technologies}, ``{Spectrum Sharing based on Shared O-RUs},''
  {O-RAN next Generation Research Group (nGRG)}, Research Report, 10 2023,
  report ID: RR-2023-05. [Online]. Available:
  \url{https://mediastorage.o-ran.org/ngrg-rr/nGRG-RR-2023-05-Spectrum\_Sharing\_with\_Shared\_O-RU-v1\_0.pdf}
\BIBentrySTDinterwordspacing

\bibitem{giordani2018tutorial}
M.~{Giordani}, M.~{Polese}, A.~{Roy}, D.~{Castor}, and M.~{Zorzi}, ``{A}
  {T}utorial on {B}eam {M}anagement for {3GPP NR} at {mmWave} {F}requencies,''
  \emph{IEEE Communications Surveys \& Tutorials}, vol.~21, no.~1, pp.
  173--196, February 2019.

\bibitem{haiPaper}
H.~Cheng, P.~Johari, M.~A. Arfaoui, F.~Periard, P.~Pietraski, G.~Zhang, and
  T.~Melodia, ``{Real-Time AI-Enabled CSI Feedback Experimentation with Open
  RAN},'' in \emph{2024 19th Wireless On-Demand Network Systems and Services
  Conference (WONS)}, 2024, pp. 121--124.

\bibitem{palama2022}
I.~Palamà, S.~Bartoletti, G.~Bianchi, and N.~B. Melazzi, ``{5G Positioning
  with SDR-based Open-source Platforms: Where do We Stand?}'' in \emph{2022
  IEEE 11th IFIP International Conference on Performance Evaluation and
  Modeling in Wireless and Wired Networks (PEMWN)}, 2022, pp. 1--6.

\bibitem{3gpp.28.552}
\BIBentryALTinterwordspacing
3GPP, ``{Management and Orchestration; 5G Performance Measurements},'' {3rd
  Generation Partnership Project (3GPP)}, Technical Specification (TS) 28.552,
  3 2022, version 17.6.0. [Online]. Available:
  \url{http://www.3gpp.org/DynaReport/28552.htm}
\BIBentrySTDinterwordspacing

\bibitem{3gpp.32.425}
\BIBentryALTinterwordspacing
------, ``{Telecommunication Management; Performance Management (PM);
  Performance Measurements Evolved Universal Terrestrial Radio Access Network
  (E-UTRAN)},'' {3rd Generation Partnership Project (3GPP)}, Technical
  Specification (TS) 32.425, 6 2021, version 17.1.0. [Online]. Available:
  \url{http://www.3gpp.org/DynaReport/32425.htm}
\BIBentrySTDinterwordspacing

\bibitem{oran-wg3-e2-sm}
{O-RAN Working Group 3}, ``{O-RAN} {N}ear-{R}eal-{T}ime {RAN} {I}ntelligent
  {C}ontroller {E2} {S}ervice {M}odel 2.00,'' ORAN-WG3.E2SM-v02.00 Technical
  Specification, 7 2021.

\bibitem{oran-wg10-OAM}
{O-RAN Working Group 10}, ``{O-RAN} {0}perations and {M}aintenance
  {A}rchitecture,'' {ORAN}-WG10O.{OAM}-Architecture-{R}004 Technical
  Specification, 7 2024.

\bibitem{oran-wg6-ORCHCL}
{O-RAN Working Group 6}, ``{C}loudification and {O}rchestration {U}se {C}ases
  and {R}equirements for {O-RAN} {V}irtualized {RAN},''
  {O-RAN-WG6}.{ORCH}-USE-CASES-R003- v11.00 Technical Specification, 7 2024.

\bibitem{hintjens2013zeromq}
P.~Hintjens, \emph{{ZeroMQ: Messaging for Many Applications}}.\hskip 1em plus
  0.5em minus 0.4em\relax O'Reilly Media, 2013.

\bibitem{zeromq-sctp}
\BIBentryALTinterwordspacing
{zhchai}, ``{libzmq not support SCTP transport },'' Github issue on the ZeroMQ
  project repository, 7 2017. [Online]. Available:
  \url{https://github.com/zeromq/libzmq/issues/2620}
\BIBentrySTDinterwordspacing

\bibitem{kaltenberger2024driving}
F.~Kaltenberger, T.~Melodia, I.~Ghauri, M.~Polese, R.~Knopp, T.~T. Nguyen,
  S.~Velumani, D.~Villa, L.~Bonati, R.~Schmidt \emph{et~al.}, ``{Driving
  Innovation in 6G Wireless Technologies: The OpenAirInterface Approach},''
  \emph{arXiv preprint arXiv:2412.13295}, 2024.

\bibitem{KALTENBERGER2020107284}
\BIBentryALTinterwordspacing
F.~Kaltenberger, A.~P. Silva, A.~Gosain, L.~Wang, and T.-T. Nguyen,
  ``{OpenAirInterface: Democratizing innovation in the 5G Era},''
  \emph{Computer Networks}, vol. 176, p. 107284, 2020. [Online]. Available:
  \url{https://www.sciencedirect.com/science/article/pii/S1389128619314410}
\BIBentrySTDinterwordspacing

\bibitem{bimo_osc_2022}
\BIBentryALTinterwordspacing
F.~A. Bimo, F.~Feliana, S.-H. Liao, C.-W. Lin, D.~F. Kinsey, J.~Li, R.~Jana,
  R.~Wright, and R.-G. Cheng, ``{OSC} {Community} {Lab}: {The} {Integration}
  {Test} {Bed} for {O}-{RAN} {Software} {Community},'' in \emph{2022 {IEEE}
  {Future} {Networks} {World} {Forum} ({FNWF})}, Oct. 2022, pp. 513--518, iSSN:
  2770-7679. [Online]. Available:
  \url{https://ieeexplore.ieee.org/document/10056724}
\BIBentrySTDinterwordspacing

\bibitem{10.1145/3485983.3494870}
\BIBentryALTinterwordspacing
R.~Schmidt, M.~Irazabal, and N.~Nikaein, ``{FlexRIC: an SDK for next-generation
  SD-RANs},'' in \emph{Proceedings of the 17th International Conference on
  Emerging Networking EXperiments and Technologies}, ser. CoNEXT '21.\hskip 1em
  plus 0.5em minus 0.4em\relax New York, NY, USA: Association for Computing
  Machinery, 2021, p. 411–425. [Online]. Available:
  \url{https://doi.org/10.1145/3485983.3494870}
\BIBentrySTDinterwordspacing

\bibitem{mundlamuri20245g}
R.~Mundlamuri, R.~Gangula, F.~Kaltenberger, and R.~Knopp, ``{5G NR} positioning
  with {OpenAirInterface}: Tools and methodologies,'' in \emph{20th Wireless
  On-Demand Network Systems and Services Conference {(WONS)}}, 2025.

\bibitem{gangula2024listen}
R.~Gangula, A.~Lacava, M.~Polese, S.~D'Oro, L.~Bonati, F.~Kaltenberger,
  P.~Johari, and T.~Melodia, ``{Listen-While-Talking: Toward dApp-based
  Real-Time Spectrum Sharing in O-RAN},'' \emph{MILCOM 2024-2024 IEEE Military
  Communications Conference (MILCOM)}, pp. 651--652, 2024.

\bibitem{3gpp.38.214}
\BIBentryALTinterwordspacing
3GPP, ``{NR}; {P}hysical layer procedures for data,'' {3rd Generation
  Partnership Project (3GPP)}, Technical Specification (TS) 38.214, 09 2024,
  version 18.4.0. [Online]. Available:
  \url{http://www.3gpp.org/DynaReport/38214.htm}
\BIBentrySTDinterwordspacing

\bibitem{bourdoux20206g}
A.~Bourdoux, A.~N. Barreto, B.~van Liempd, C.~de~Lima, D.~Dardari, D.~Belot,
  E.-S. Lohan, G.~Seco-Granados, H.~Sarieddeen, H.~Wymeersch \emph{et~al.},
  ``{6G White Paper on Localization and Sensing},'' \emph{arXiv preprint
  arXiv:2006.01779}, 2020.

\bibitem{Nessa2020mlpos}
A.~Nessa, B.~Adhikari, F.~Hussain, and X.~N. Fernando, ``{A Survey of Machine
  Learning for Indoor Positioning},'' \emph{IEEE Access}, vol.~8, pp.
  214\,945--214\,965, 2020.

\bibitem{xinrong2004}
X.~Li and K.~Pahlavan, ``{S}uper-{R}esolution {TOA} {E}stimation {W}ith
  {D}iversity for {I}ndoor {G}eolocation,'' \emph{IEEE Transactions on Wireless
  Communications}, vol.~3, no.~1, pp. 224--234, 2004.

\bibitem{gangula2024round}
R.~Gangula, T.~Melodia, R.~Mundlamuri, and F.~Kaltenberger, ``{Round Trip Time
  Estimation Utilizing Cyclic Shift of Uplink Reference Signal},'' \emph{arXiv
  preprint arXiv:2410.04528}, 2024.

\bibitem{sc_mod}
SignalCraft,
  \url{https://www.signalcraft.com/products/test-measurement/microwave-systems/sc2430/}.

\bibitem{villa2024x5g}
D.~Villa, I.~Khan, F.~Kaltenberger, N.~Hedberg, R.~S. da~Silva, S.~Maxenti,
  L.~Bonati, A.~Kelkar, C.~Dick, E.~Baena \emph{et~al.}, ``{X5G: An Open,
  Programmable, Multi-vendor, End-to-end, Private 5G O-RAN Testbed with NVIDIA
  ARC and OpenAirInterface},'' \emph{arXiv preprint arXiv:2406.15935}, 2024.

\end{thebibliography}
\end{document}